\documentclass[3p,times]{elsarticle}
\usepackage{textcomp}
\usepackage{enumitem}
\newlist{todolist}{itemize}{2}
\setlist[todolist]{label=$\square$}
\usepackage{pifont}
\usepackage{amssymb}
\usepackage{amsthm}
\usepackage{amsmath}

\usepackage[figuresright]{rotating}

\usepackage{subfig}

\usepackage{graphicx}% Include figure files
\usepackage{dcolumn}% Align table columns on decimal point
\usepackage{bm}% bold math
\usepackage[colorlinks = true,
linkcolor = blue,
urlcolor  = blue,
citecolor = blue,
anchorcolor = blue]{hyperref}% add hypertext capabilities
\usepackage{url}
\usepackage{physics}

\begin{document}
	
	\begin{frontmatter}
%		\dochead{}
		
		%		\title{From Flat Bands to Energy Bursts: \\In-band supratransmissions in diamond and stub lattices}
		\title{In-band supratransmission in nonlinear flat band lattices}
		
		
		\author[au1,au1b]{R.\ Kusdiantara\corref{cor1}}
		\ead{rudy\_kusdiantara@itb.ac.id}	
		\author[au1]{M.\ Wijaya}
		\author[au1]{M.\ F.\ Adhari}
		%		\author[au3]{O.B.\ Kirikchi}%
		%		\author[au4]{N.\ Karjanto\corref{cor1}}%
		%		\ead{natanael@skku.edu}
		\author[au5]{H.\ Susanto}
		%		\ead{hadi.susanto@ku.ac.ae}
		%		\author[au2]{A. R. Champneys}	
		
		\address[au1]{Industrial and Financial Mathematics Research Group, Institut Teknologi Bandung, Jl.\ Ganesha No.\ 10, Bandung, 40132, Indonesia}
		\address[au1b]{Centre of Mathematical Modelling and Simulation, Institut Teknologi Bandung, Jl.\ Ganesha No.\ 10, Bandung, 40132, Indonesia}
		\cortext[cor1]{Corresponding author}
		%		\address[au2]{Department of Mathematics and Statistics, College of Science, Taif University, P.O.Box 11099, Taif 21944, Saudi Arabia}
		%		\address[au3]{Department of Computing, Goldsmiths, University of London, New Cross, London, SE14 6AD, United Kingdom}
		%		\address[au4]{Department of Mathematics, University College, Natural Science Campus, Sungkyunkwan University, 2066 Seobu-ro, Jangan-gu, Suwon, Gyeonggi-do, 16419, Republic of Korea}
		\address[au5]{Department of Mathematics, Khalifa University, PO Box 127788, Abu Dhabi, United Arab Emirates}
		%		\address[au3]{Department of Mathematics, Faculty of Mathematics and Natural Sciences, Universitas Indonesia,\\ Gedung D Lt.\ 2 FMIPA Kampus UI Depok, 16424, Indonesia}
		%		\cortext[cor1]{Corresponding author}
		%		\address[au2]{Department of Engineering Mathematics, University of Bristol, Bristol BS8 1UB, United Kingdom}
		%\affiliation{%
			%	Department of Mathematical Sciences, University of Essex, Wivenhoe Park, Colchester CO4 3SQ, United Kingdom
			%}%

		\begin{abstract}
			Studying wave propagation in nonlinear discrete systems is essential for understanding energy transfer in lattices. While linear systems prohibit wave propagation within the natural band gap, nonlinear systems exhibit {supratransmission}, enabling energy transfer above a critical driving amplitude. This work investigates novel \emph{in-band supratransmissions} for waves with frequencies in a \emph{flat} or \emph{nearly flat} linear band. Flat bands, characterized by zero group velocity and localized modes due to destructive interference, provide an ideal framework for studying wave confinement and energy dynamics. In-band supratransmission originates from a bifurcation of evanescent waves at the flat band frequency. Using nonlinear \emph{diamond} and \emph{stub} lattices as model systems, we explore how lattice topology, nonlinearity, and driving amplitude affect supratransmission. 
			Through bifurcation analysis, stability evaluations, and time-dependent simulations, we examine the transition from energy localization to supratransmission.
		\end{abstract}
		

	\end{frontmatter}
	
	% \linenumbers
	
	\section{Introduction}
	
	Periodic systems, such as crystal lattices, are characterized by a repeating arrangement of atoms or unit cells, creating a periodic potential that influences the behavior of waves—whether electronic, electromagnetic, or acoustic \cite{Kittel2005}. The propagation of waves in such systems is described by the dispersion relation, which connects the wavevector to the frequency or energy of the waves. This relation provides critical insights into the system's dynamics, including the group velocity of wave packets (determined by the slope of the dispersion relation) and the formation of energy bands, known as dispersive bands, separated by frequency ranges where wave propagation is forbidden (band gaps) \cite{Ashcroft1976}. For example, insulators exhibit wide band gaps that prevent electron excitation, while semiconductors have smaller gaps, enabling controlled conductivity under specific conditions and making them indispensable for modern electronics \cite{Ashcroft1976}. Band gaps also play a pivotal role in photonic and phononic crystals. In photonic crystals, they control the propagation of electromagnetic waves, facilitating the design of optical filters and waveguides \cite{Joannopoulos2008}. Similarly, in phononic crystals, band gaps regulate acoustic wave propagation, enabling innovations in vibration isolation and soundproofing technologies \cite{deymier2013acoustic}.
	
{When nonlinear effects are introduced in discrete systems, the characteristics of waves with frequencies within band gaps can be significantly altered, leading to nontrivial phenomena such as soliton formations \cite{Flach2008,trillo2013spatial}, i.e., localized wave packets arising from the interplay between nonlinearity and dispersion, and, notably, supratransmission. Supratransmission is a nonlinear transmission mechanism in which energy, otherwise blocked by a forbidden band gap in the linear regime, can propagate through the medium once the input amplitude exceeds a critical threshold \cite{geniet2003nonlinear,leon2003nonlinear,leon2004gap}. Gap solitons mediate this energy transfer. Supratransmission has been experimentally verified in electrical transmission systems, as documented in studies by \cite{bodo2010klein,koon2014experimental,tao2012experimental}. Similar effects have been identified in finite-length monoatomic granular chains subjected to harmonic driving forces, as explored by \cite{lydon2015nonlinear}. Furthermore, the ability to regulate supratransmission has been achieved in mechanical metastructures featuring intricate folding patterns, including Miura origami configurations \cite{zhang2020programmable} and Kresling origami architectures \cite{wang2023highly}.}
	
	A recent study \cite{susanto2023surge} reported \emph{in-band supratransmission}, where nonlinear energy transfer occurs in discrete systems with frequencies within a flat or nearly flat dispersive band. Flat bands, characterized by constant frequencies throughout wavenumbers, arise from specific symmetries or destructive interference, often in geometrically frustrated or highly symmetric systems \cite{leykam2018artificial}. These bands exhibit vanishing group velocity, enabling phenomena such as slow light in photonic systems \cite{settle2007flatband} or strongly correlated electron behavior \cite{derzhko2015strongly}. Unlike standard supratransmission, which relies on gap solitons for energy transfer \cite{leon2003nonlinear,leon2004gap,macias2018supratransmission,susanto2008boundary,tchameu2016nonlinear,susanto2008calculated,motcheyo2017homoclinic, togueu2019supratransmission, kuitche2022supratransmission,motcheyo2024nonlinear}, in-band supratransmission is mediated by nonlinear plane waves, highlighting a unique feature of nonlinear dynamics in flat band lattices. Numerical experiments on in-band supratransmission were reported in \cite{bebikhov2024discrete}. See also \cite{motcheyo2023nonlinear} for in-band supratransmission with different setups. 
	
	This work investigates in-band supratransmission further, focusing on the role of lattice topology, i.e., the geometric arrangement of sites and their connectivity. Topology dictates the formation of energy bands, localized states, and wave propagation robustness against disorder and perturbations \cite{kartashov2009soliton,leykam2015lattice}. Herein, we are interested in its influences on nonlinear dynamics. We examine two prototypical models: the one-dimensional (1D) diamond and stub lattices \cite{huda2020designer,real2017flat,caceres2022experimental}. The diamond lattice, a 1D chain with a two-atom basis, is often regarded as the 1D analog of the 2D honeycomb lattice \cite{fefferman2012honeycomb,sinelnik2017optical}. It exhibits a flat band at the center of the Brillouin zone, associated with compact localized states where wave energy is confined to a finite number of lattice sites. This flat band enhances phenomena like ferromagnetism and superconductivity due to the high density of states at the flat band energy \cite{peri2021fragile}. In contrast, the stub lattice, a decorated 1D chain with an additional side atom (stub), introduces asymmetry that disrupts translational symmetry \cite{huda2020designer}. This disruption can create additional flat bands, depending on hopping parameters and relative energies of the stub and chain atoms. The localization patterns in the stub lattice differ from those in the diamond lattice due to its asymmetric structure. In our setup, the stub lattice is derived from the diamond lattice by eliminating one connectivity, providing a platform to explore the effect of lattice topology on in-band nonlinear supratransmission.
	
{The novelties of this work lie in extending the concept of in-band supratransmission from one of the authors' prior study \cite{susanto2023surge} on the perturbed sawtooth lattice to the diamond and stub lattices. Firstly, \cite{susanto2023surge} demonstrated that a localized mode can arise from an edge drive with an amplitude below the transmission threshold. Here, we observe similar behavior in both the diamond and stub lattices, indicating that edge-driven localized modes are not unique to a specific geometry but are instead a more general feature across different flatband systems. Secondly, while localized modes are typically confined to perfectly flat bands, our results show that a supratransmission threshold also exists for nearly flat bands. Although the precise origin of this phenomenon remains unclear, it is most likely linked to nonlinear effects, which can sustain or suppress transmission even when the band is not perfectly flat. Lastly, the influence of topologically localized states, previously investigated in \cite{susanto2023surge}, can also be examined in the present lattices. In the diamond lattice in particular, the system offers an additional level of tunability: not only can the location and existence of the flat band be adjusted, but the emergence of topologically localized states can also be controlled. This combined ability to manipulate both flatband and topological features enables a deeper exploration of their impact on edge-driven dynamics, nonlinear wave propagation, and supratransmission.}
	
	%	\textcolor{red}{		Recent studies on supratransmission in periodic nonlinear systems have primarily focused on dispersive lattices, where energy transfer occurs more uniformly. In contrast, our study highlights the unique role of flat-band geometry in nonlinear supratransmission. Flat-band systems, characterized by interference-based localization and compact localized modes (CLMs), enable energy confinement that is sensitive to nonlinear interactions. The activation of transport through nonlinear coupling in these systems is distinct from traditional systems, where energy transfer is more uniform. These features of flat-band geometry--particularly compactness and interference-based localization--allow supratransmission to occur at lower driving amplitudes \cite{leykam2018artificial}, emphasizing the novel mechanisms at play in our study. This approach provides new insights into how lattice topology and nonlinearity can combine to drive energy transport in a way that is not typically observed in conventional systems \cite{geniet2003nonlinear}. Photonic crystals and optical lattices, which are established experimental platforms, offer a way to simulate the discrete systems studied here and can help validate the theoretical findings of nonlinear supratransmission in flat-band lattices.	}
	
	This paper is organized as follows. Section~\ref{sec:model} describes the mathematical models of the diamond and stub lattices, including the governing equations, coupling configurations, and flat-band properties. Because it is known that supratransmission is strongly related to evanescent waves ceasing to exist \cite{susanto2008boundary} or becoming unstable \cite{susanto2023surge}, we begin our study by considering steady-state solutions, focusing on bifurcation diagrams and the stability of localized modes. This is presented in  Section~\ref{sec:independent}. Section~\ref{sec:dependent} discusses time-dependent dynamics, highlighting the interplay between nonlinearity and lattice topology in supratransmission. Finally, Section~\ref{sec:conc} presents conclusions, summarizing the key findings and their implications for nonlinear discrete systems.

	\section{Mathematical Model}\label{sec:model}
	%	\begin{todolist}
		%		\item[\done] Equations of diamond \& Stub
		%		\item[\done] Ilustrations diamond and stub
		%		\item[\done] Dispersion relation $k$ vs $\Omega$
		%		\item[\done] Frequency bands vs coupling
		%	\end{todolist}
	
	\begin{figure}[tbhp!]
		\centering
		%\subfloat[Diamond Lattice]
		{\includegraphics[scale=0.7]{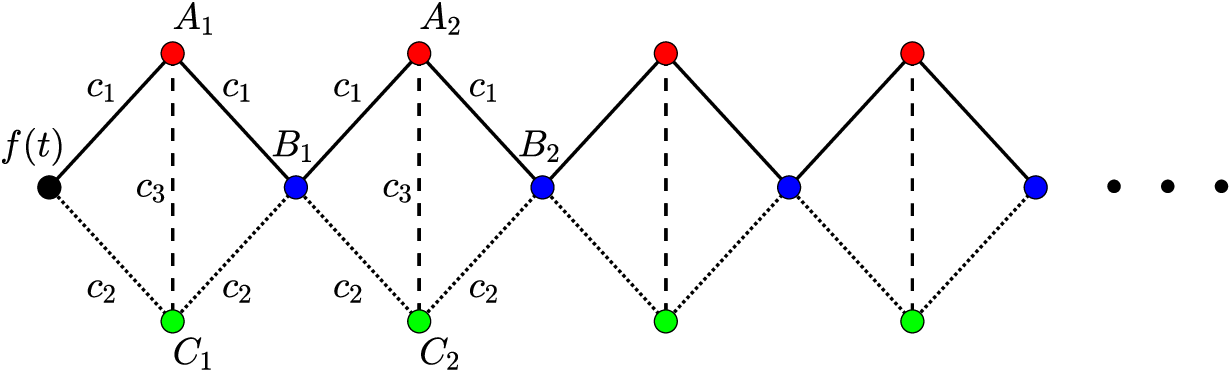}}\\%\label{subfig:Diamond_display}}\\
	%	\subfloat[Stub Lattice]{\includegraphics[scale=0.7]{Stub_display}\label{subfig:Stub_display}}
	\caption{Schematic representation of an edge-driven one-dimensional diamond lattice. %(a) Diamond lattice, characterized by couplings \(c_1 \neq 0\), \(c_2\), and \(c_3\), allowing more connectivity and wide propagation bands. (b) 
		A stub lattice is when \(c_1 = 0\), leading to reduced direct connectivity between sites $A_n$ and $B_n$.}
	\label{fig:diamond_stub}
\end{figure}
The diamond and stub lattices, shown in Fig.~\ref{fig:diamond_stub}, are the focus of this study. The governing equations describing the systems are:
\begin{equation}
	%		\begin{array}{rcl}
		%			i \dot{A}_n &=& c_1 \left(B_{n-1} + B_n - 2A_n\right) + c_3\left(C_n - A_n\right) + \gamma |A_n|^2A_n, \\
		%			i \dot{B}_n &=& c_1 \left(A_n + A_{n+1} - 2B_n\right) + c_2\left(C_n + C_{n+1} - 2B_n\right) + \gamma |B_n|^2B_n, \\
		%			i \dot{C}_n &=& c_2 \left(B_{n-1} + B_n - 2C_n\right) + c_3\left(A_n - C_n\right) + \gamma |C_n|^2C_n,
		%		\end{array}
	\begin{array}{rcl}
		i \dot{A}_n &=& c_1 \left(\Delta_b B_{n} + B_n - 2A_n\right) + c_3\left(C_n - A_n\right) + \gamma |A_n|^2A_n, \\
		i \dot{B}_n &=& c_1 \left(A_n + \Delta_f A_{n} - 2B_n\right) + c_2\left(C_n + \Delta_f C_{n} - 2B_n\right) + \gamma |B_n|^2B_n, \\
		i \dot{C}_n &=& c_2 \left(\Delta_b B_{n} + B_n - 2C_n\right) + c_3\left(A_n - C_n\right) + \gamma |C_n|^2C_n,
	\end{array}
	\label{eq:supra_ori}
\end{equation}
{
	$A_n$, $B_n$, and $C_n$ denote the complex wave amplitudes at the $n$-th unit cell for the three sublattices of the system. The overdot represents differentiation with respect to the temporal variable $t$, or equivalently, the propagation coordinate in the context of nonlinear optics. The operators $\Delta_{f} \square_n = \square_{n+1}$ and $\Delta_{b} \square_n = \square_{n-1}$ correspond to forward and backward lattice shifts, respectively. Here, $n \in \mathbb{Z}^+$, $\gamma$ is the nonlinearity coefficient, and $c_1$, $c_2$, and $c_3$ are the coupling strengths between adjacent sites. The nonlinear term $|\square_n|^2 \square_n$ models Kerr-type self-interaction within each sublattice, introducing cubic nonlinearity into the system. The sign of the coefficient $\gamma$ determines the nature of the nonlinearity: a positive value ($\gamma > 0$) corresponds to focusing nonlinearity, where wave packets attract and can form tightly localized structures; a negative value ($\gamma < 0$) represents defocusing nonlinearity, leading to repulsive interactions and more delocalized wave propagation. As we will show below, these two regimes show different dynamical characteristics %critically influence the formation, stability, and dynamics of localized modes 
	in the lattice. }
The system is driven at its left boundary with 
\begin{equation}
	B_0 = f(t) = F(t)e^{-i \Omega t}. \label{B0}
\end{equation}
The driving amplitude \(F(t)\) is turned on gradually as \(F(t) = f_0(1 - e^{-t/\tau})\), with \(\tau \gg 1\) to avoid initial shock. 

The diamond lattice is defined by \(c_1 \neq 0\), which allows strong connectivity between sites. This will be shown to have wide frequency bands and expectedly large energy transfer. On the other hand, the stub lattice is characterized by \(c_1 = 0\), resulting in reduced connectivity and narrower dispersive bands, as we will show below.
%\textcolor{red}{Topology also plays a critical role in determining the band structure, energy localization, and wave propagation behavior in nonlinear lattice systems. The connectivity of the lattice dictates the formation of dispersive and flat bands, which in turn affect the stability of localized modes and the onset of supratransmission. For example, in the diamond lattice, with its higher connectivity, we observe wider frequency bands and more efficient energy transfer, while the stub lattice, with reduced connectivity, exhibits narrower dispersive bands and higher thresholds for supratransmission. These differences underscore the influence of lattice topology on the nonlinear dynamics of the system.}

The steady state of the system Eq.\ \eqref{eq:supra_ori} is obtained by considering solutions of the form:
\begin{equation}
	\left(
	\begin{array}{c}
		A_n(t) \\ B_n(t) \\ C_n(t)
	\end{array} \right) =
	\left(
	\begin{array}{c}
		\tilde{A}_n \\ \tilde{B}_n \\ \tilde{C}_n
	\end{array} \right) e^{-i \Omega t}.
\end{equation}
Substituting it into Eq.\ \eqref{eq:supra_ori} yields:
\begin{equation}
	%		\begin{array}{rcl}
		%			\Omega \tilde{A}_n &=& c_1 \left(\tilde{B}_{n-1} + \tilde{B}_n - 2\tilde{A}_n\right) + c_3 \left(\tilde{C}_n - \tilde{A}_n\right) + \gamma \tilde{A}_n^3, \\
		%			\Omega \tilde{B}_n &=& c_1 \left(\tilde{A}_n + \tilde{A}_{n+1} - 2\tilde{B}_n\right) + c_2 \left(\tilde{C}_n + \tilde{C}_{n+1} - 2\tilde{B}_n\right) + \gamma \tilde{B}_n^3, \\
		%			\Omega \tilde{C}_n &=& c_2 \left(\tilde{B}_{n-1} + \tilde{B}_n - 2\tilde{C}_n\right) + c_3 \left(\tilde{A}_n - \tilde{C}_n\right) + \gamma \tilde{C}_n^3.
		%		\end{array}
	\begin{array}{rcl}
		\Omega \tilde{A}_n &=& c_1 \left(\Delta_b\tilde{B}_{n} + \tilde{B}_n - 2\tilde{A}_n\right) + c_3 \left(\tilde{C}_n - \tilde{A}_n\right) + \gamma \tilde{A}_n^3, \\
		\Omega \tilde{B}_n &=& c_1 \left(\tilde{A}_n + \Delta_f \tilde{A}_{n} - 2\tilde{B}_n\right) + c_2 \left(\tilde{C}_n + \Delta_f \tilde{C}_{n} - 2\tilde{B}_n\right) + \gamma \tilde{B}_n^3, \\
		\Omega \tilde{C}_n &=& c_2 \left(\Delta_b\tilde{B}_{n} + \tilde{B}_n - 2\tilde{C}_n\right) + c_3 \left(\tilde{A}_n - \tilde{C}_n\right) + \gamma \tilde{C}_n^3.
	\end{array}
	\label{eq:supra_TI}
\end{equation}
To analyze the system's frequency bands, we take the ansatz:
\begin{equation}
	\left(
	\begin{array}{c}
		\tilde{A}_n \\ \tilde{B}_n \\ \tilde{C}_n
	\end{array} \right) =
	\left(
	\begin{array}{c}
		\epsilon_1 \\ \epsilon_2 \\ \epsilon_3
	\end{array} \right) e^{i k n},
\end{equation}
and linearize around \( \epsilon_1 = \epsilon_2 = \epsilon_3 = 0 \). This results in the eigenvalue problem:
\begin{equation}
	\Omega \left(
	\begin{array}{c}
		\epsilon_1 \\ \epsilon_2 \\ \epsilon_3
	\end{array} \right) =
	M \left(
	\begin{array}{c}
		\epsilon_1 \\ \epsilon_2 \\ \epsilon_3
	\end{array} \right),\label{eig}
\end{equation}
where the matrix \( M \) is given by:
\begin{equation}
	M = \left(
	\begin{array}{ccc}
		-2c_1 - c_3 & c_1 (e^{-i k} + 1) & c_3 \\
		c_1 (e^{i k} + 1) & -2(c_1 + c_2) & c_2 (e^{i k} + 1) \\
		c_3 & c_2 (e^{-i k} + 1) & -2c_2 - c_3
	\end{array} \right).
\end{equation}
Solving the eigenvalue problem \eqref{eig} will yield the dispersion relation \(\Omega = \Omega(k;c_1,c_2,c_3)\), which we do not write here for simplicity. 

\begin{figure}[tbhp!]
	\centering
	\subfloat[Diamond Lattice for $c_1=c_2=2$, $c_3=1$]{\includegraphics[scale=0.39]{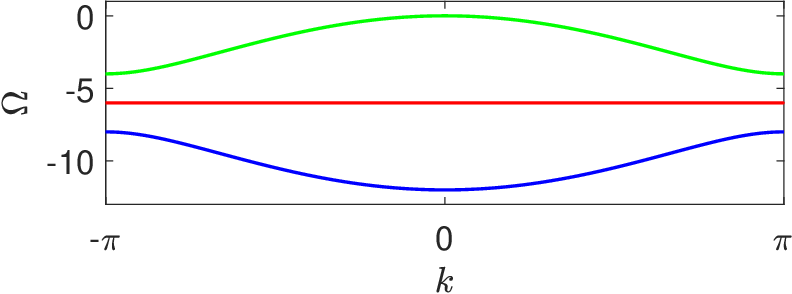}\label{subfig:Diamond_disper}}\quad
	\subfloat[Diamond Lattice for $c_1=c_2=1$, $c_3=1.5$]{\includegraphics[scale=0.39]{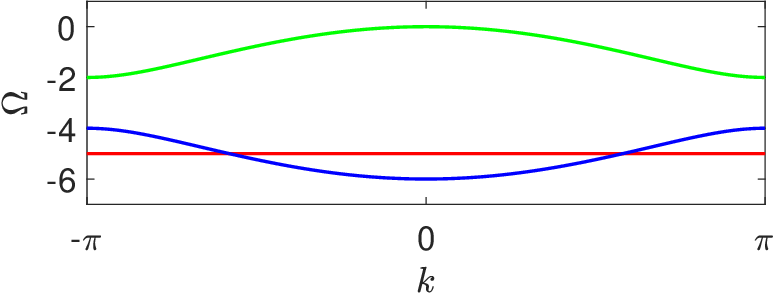}\label{subfig:Diamond_disper_inner}}\quad 
	\subfloat[Stub Lattice for $c_1=0,\, c_2=1$, $c_3=2$]{\includegraphics[scale=0.39]{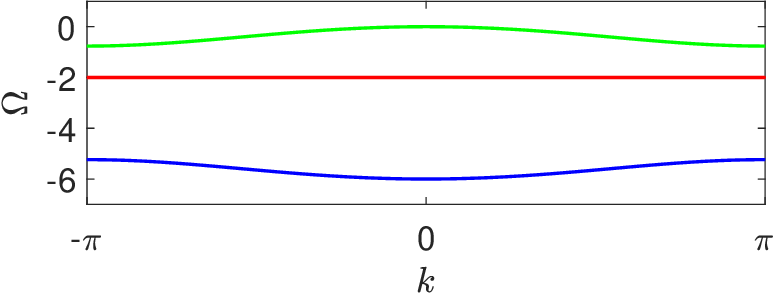}\label{subfig:Stub_disper}}\quad
	
	%		\caption{Diamond and Stub lattices}
	\caption{Dispersion relations \(\Omega(k)\) for the diamond and stub lattices {in the the first Brillouin zone $k\in [-\pi,\pi]$}. Three bands are observed, i.e., a flat band (red line) representing localized modes and two dispersive bands (green and blue lines) indicating propagating waves.
	{ (a,b) Diamond lattice: \(c_1 = c_2 = 2,\ c_3 = 1\) and \(c_1 = c_2 = 1,\ c_3 = 1.5\), respectively} (c) Stub lattice: \(c_1 = 0\), \(c_2 = 1\), and \(c_3 = 2\). The absence of coupling \(c_1\) results in narrower dispersive bands while retaining the flat band.
	}		
	\label{fig:diamond_stub_disper}
\end{figure}

Figure~\ref{fig:diamond_stub_disper} illustrates the dispersion relations \(\Omega(k)\) for the diamond and stub lattices for fixed coupling parameters. Three bands are observed: a flat band (red line) representing localized modes with zero energy propagation and two dispersive bands (green and blue lines) corresponding to propagating waves. The dispersive bands' frequencies reflect the lattice's translational symmetry. 
{In the case of the diamond lattice, the coupling configuration allows tunability of the flat band's position relative to the dispersive bands, as demonstrated in panel~(b). Here, we aim to observe the dynamic differences compared to the condition in panel (a).}
Compared to the diamond lattice, the stub lattice (\(c_1 = 0\)) exhibits a narrower band structure due to reduced connectivity. While the flat band persists, the dispersive bands shrink in bandwidth, highlighting how the absence of \(c_1\) restricts wave interactions. These features demonstrate the critical role of the lattice structures and coupling parameters in shaping wave dynamics. 

%Recent work by Leykam~\emph{et al.} \cite{leykam2018artificialartificial,leykam2024flat} defines two flat‐band classes: \emph{Singular}, which touch dispersive bands at isolated $k$-points (allowing hybridization with extended states), and \emph{Isolated}, which lie fully within a bandgap (remaining spectrally separated).
%In both our diamond ($c_1 = c_2 = 2$, $c_3 = 1$) and stub ($c_1 = 0$, $c_2 = 2$, $c_3 = 1$) lattices the flat band never intersects the dispersive branches, confirming its isolated character and the clear separation from extended modes.

\begin{figure}[tbhp!]
	\centering
	\subfloat[Diamond Lattice]{\includegraphics[scale=0.5]{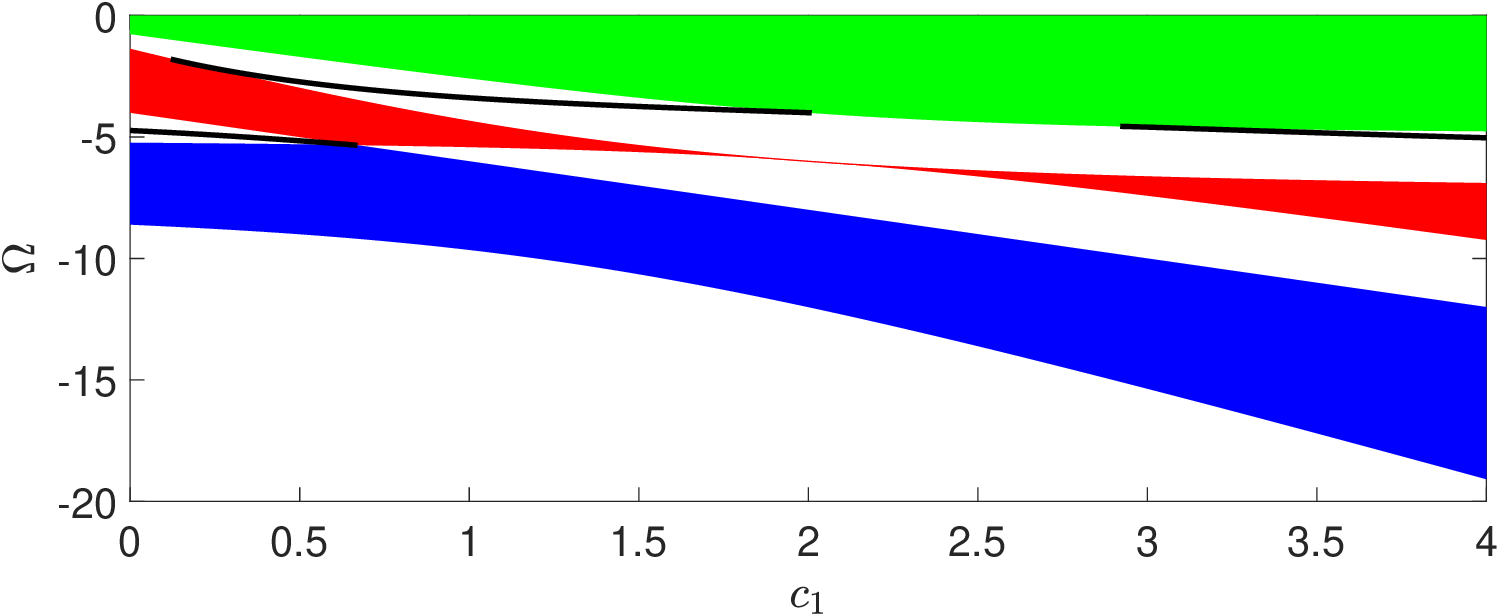}\label{subfig:diamond_freq_band}}\\
	\subfloat[Diamond Lattice]{\includegraphics[scale=0.5]{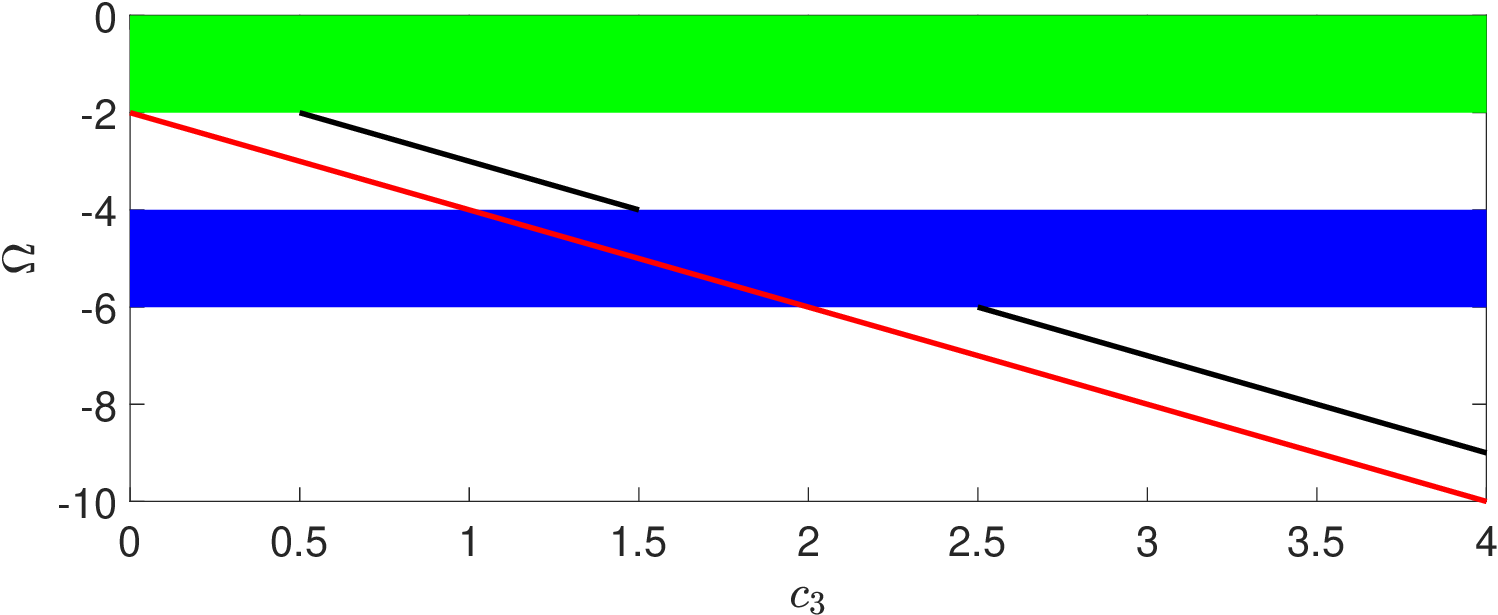}\label{subfig:diamond_freq_band_inner}}\\
	\subfloat[Stub Lattice]{\includegraphics[scale=0.5]{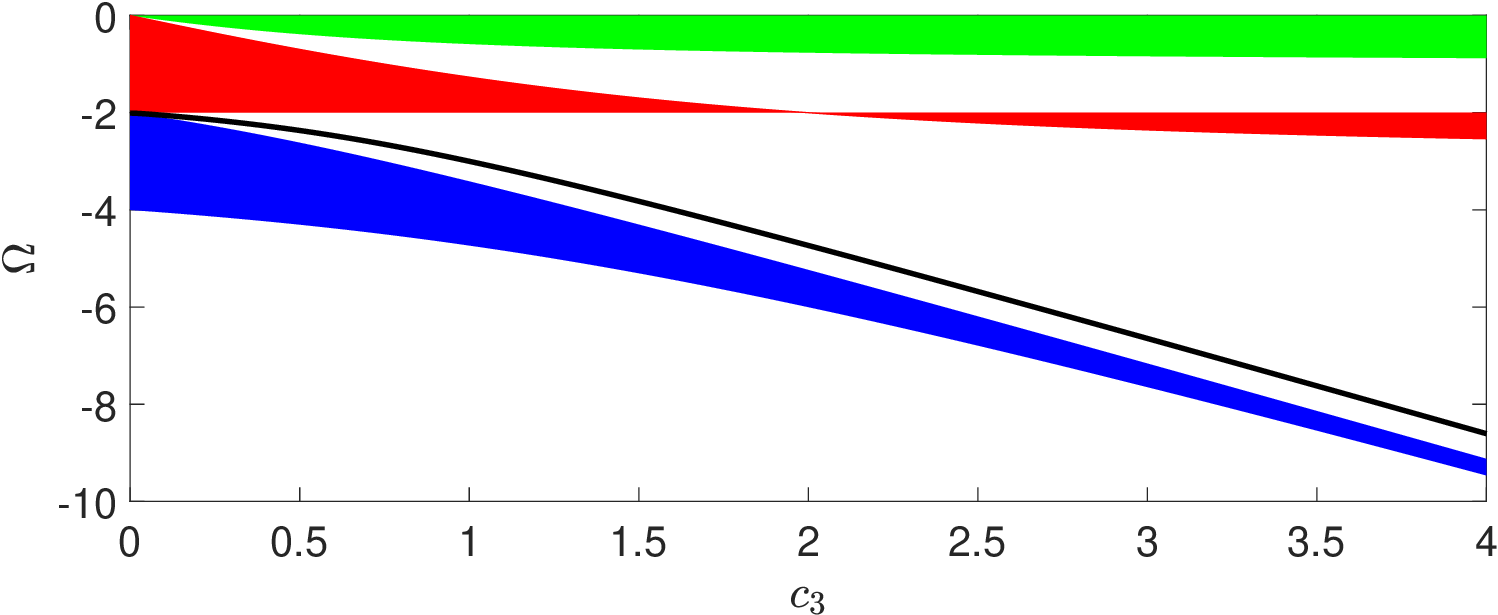}\label{subfig:stub_freq_band}}	
	\caption{(Color online) Frequency band structures for the diamond and stub lattices.
		(a) Diamond lattice with fixed coupling parameters \(c_2 = 2\) and \(c_3 = 1\), showing a wider band structure due to stronger inter-site coupling.
		(b) Diamond lattice with \(c_1 = c_2 = 1\), where \(c_3\) is varied. The flat band (red line) persists across most values of \(c_3\). Within the interval \(c_3 \in [1, 2]\), it exists inside the continuous spectrum. The dispersive bands remain unchanged.
		(c) Stub lattice with fixed \(c_2 = 1\) and varying \(c_3\), illustrating narrower frequency bands due to reduced connectivity. The flat band appears in both lattices (highlighted in red), while the continuous dispersive bands (green and blue regions) represent propagating modes. Additional discrete spectral branches (black thick lines) correspond to topological states localized at the boundary.}
	\label{fig:diamond_stub_freq}
\end{figure}

The frequency band structures for the diamond and stub lattices are shown in Fig.~\ref{fig:diamond_stub_freq}. For the diamond lattice (Fig.~\ref{subfig:diamond_freq_band}), we fix the coupling coefficients \(c_2\) and \(c_3\), and vary \(c_1\). The flat band (highlighted in red) emerges distinctly at \(c_1 = 1\).

Additionally, we fix \(c_1 = c_2 = 1\) and vary \(c_3\). In this scenario, the flat band—marked by the red dashed line—persists across the parameter range and enters the continuous spectrum when \(c_3 \in [1,2]\), indicating a transition from isolated to embedded flat-band behavior.

For the stub lattice (Fig.~\ref{subfig:stub_freq_band}), we fix \(c_2\) and vary \(c_3\). The flat band appears at \(c_3 = 2\), and the dispersive bands become progressively narrower as \(c_3\) increases. By rescaling the system, a similar narrowing effect can be observed when \(c_3\) is held fixed and \(c_2\) is decreased. 

%\textcolor{red}{Leykam \emph{et al.}~\cite{leykam2018artificialartificial,leykam2024flat} classify flat bands as \emph{singular}-touching dispersive bands at isolated $k$-points--and \emph{isolated}--lying entirely within a gap.  In both our diamond ($c_1=c_2=2$, $c_3=1$) and stub ($c_1=0$, $c_2=2$, $c_3=1$) lattices the flat band never intersects the dispersive branches, confirming its isolated character.}

{Flat bands can be broadly classified based on their energetic relation to adjacent dispersive bands. In one category, flat bands intersect or touch dispersive bands at specific points in momentum space, forming so-called ``touching'' flat bands \cite{graf2021designing,rhim2019classification}. These intersections often lead to singularities in the density of states and are typically associated with nontrivial topological features or symmetry-protected degeneracies. In contrast, ``isolated'' flat bands are energetically separated from all dispersive bands by a finite gap \cite{leykam2024flat}. This isolation suppresses hybridization with propagating modes and enhances the localization of wavefunctions, which can promote phenomena such as compact localized states (CLSs) \cite{maimaiti2017compact,tang2020photonic}. The presence or absence of a gap significantly influences the dynamical and transport properties of the system, determining whether energy remains confined or can leak into extended modes.
	
	The band structures presented in Fig.~\ref{fig:diamond_stub_freq} clearly demonstrate that the diamond lattice exhibits tunable spectral features: the flat band can either be isolated, touch, or be embedded in the dispersive bands, depending on the system parameters. In contrast, the stub lattice consistently supports an isolated flat band. We will demonstrate that supratransmission behavior is closely linked to these spectral conditions of the lattice. In systems with touching or embedded flat bands, supratransmission can occur without a threshold amplitude, while in systems with isolated flat bands, a finite critical amplitude is typically required to initiate energy transmission~\cite{susanto2023surge}. 
	
	While the dispersion bands are obtained analytically under the assumption of an infinite domain, we also numerically solve the eigenvalue problem \eqref{eig} on a finite domain. In this setting, boundary conditions are imposed as \(\tilde{B}_0 = 0\) on the left and \(\tilde{A}_N = \tilde{C}_N = 0\) on the right, where \(N\) denotes the size of the computational grid. The resulting spectrum is shown in Fig.~\ref{fig:diamond_stub_freq}. In addition to eigenvalues lying within the analytically derived continuous bands, we observe isolated eigenvalues (plotted as thick black curves) that fall outside the bulk spectrum. These correspond to topologically protected edge states, i.e., localized modes that arise due to the finite geometry and boundary conditions, and are robust to perturbations.
	
	Notably, the stub lattice always hosts a topologically protected edge state, shown by the localized spectrum, which can be absent in the diamond lattice under the same conditions. We will show that the presence of a topological edge state in the stub lattice gives rise to a beating phenomenon, resulting from the interaction between the driving frequency and the natural frequency of the edge state. }

\section{Time-independent solutions}\label{sec:independent}

\begin{figure}[tbhp!]
	\centering
	\subfloat[Diamond Lattice Focusing]{\includegraphics[scale=0.58]{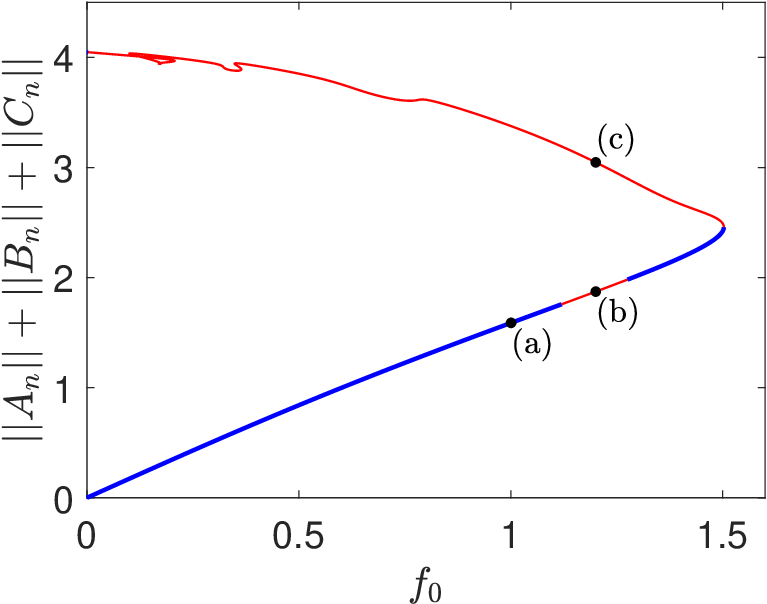}\label{subfig:bifur_diamond_focusing}}\qquad
	\subfloat[Diamond Lattice Defocusing]{\includegraphics[scale=0.58]{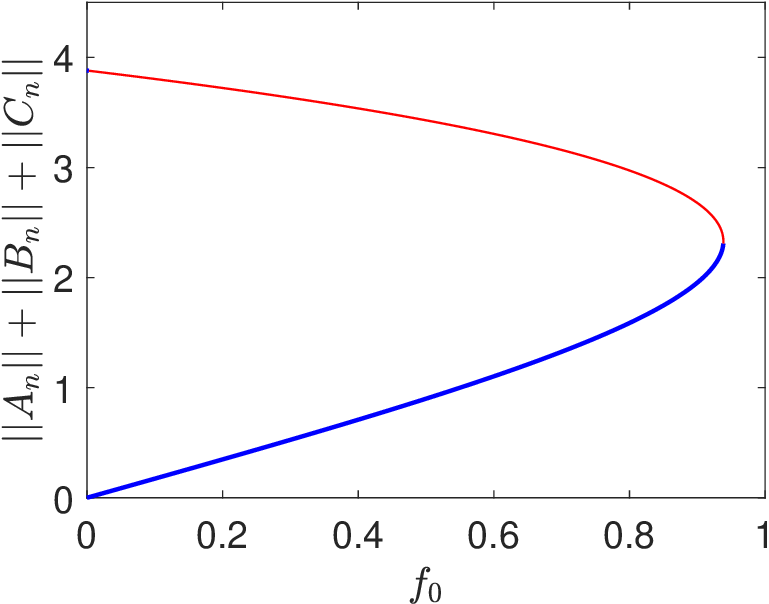}\label{subfig:bifur_diamond_defocusing}}\\
	\subfloat[Stub Lattice Focusing]{\includegraphics[scale=0.58]{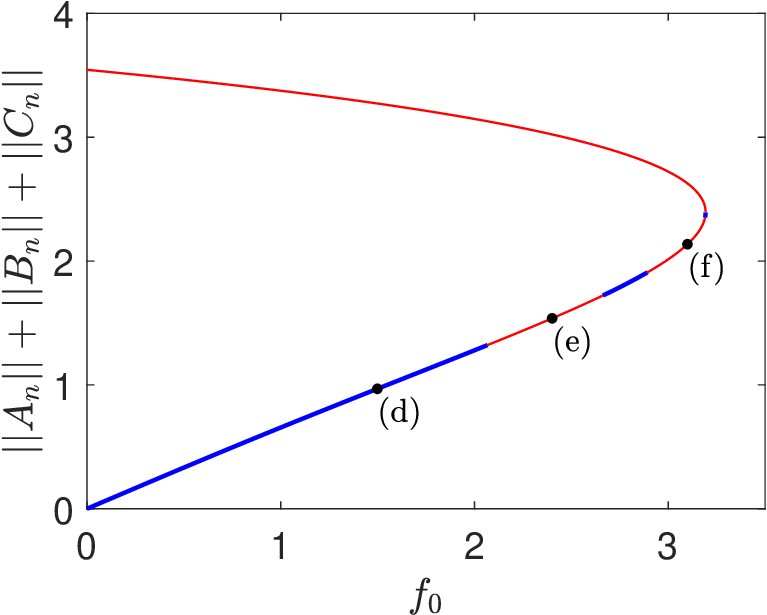}\label{subfig:bifur_stub_focusing}}\qquad
	\subfloat[Stub Lattice Defocusing]{\includegraphics[scale=0.58]{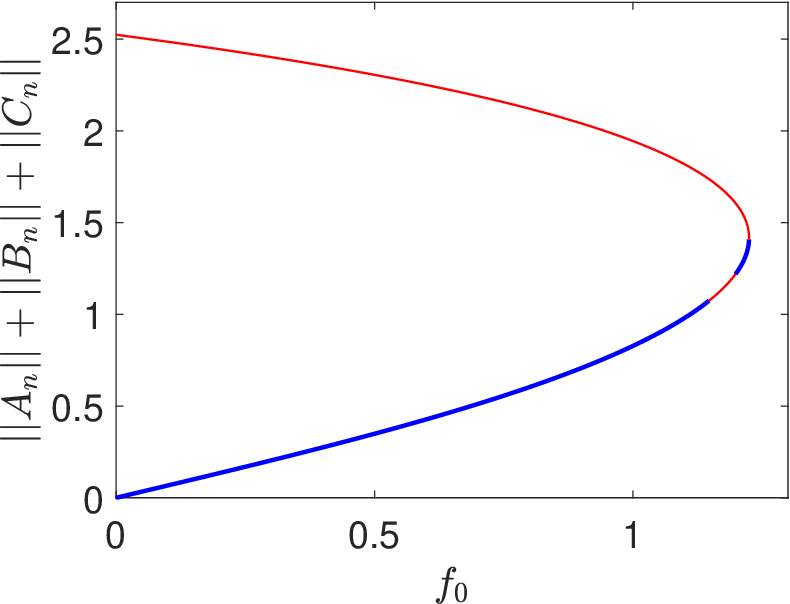}\label{subfig:bifur_stub_defocusing}}
	\caption{Bifurcation diagrams for diamond and stub lattices under focusing and defocusing nonlinearity with parameters at the flat band. In all the panels, there is a turning point for decaying solutions. %(a) Diamond lattice focusing: stable branches (blue) transition to unstable branches (red) before the turning point. (b) Diamond lattice defocusing: the lower branch is always stable (c) Stub lattice focusing: longer range of drive for decaying solutions compared to the diamond lattice, with a similar transition to instability. (d) Stub lattice defocusing: mostly stable solutions with small instability regions along the lower branch. 
		{They are $f_0=1.5026;0.9385;3.1928;1.2246$ for panels (a)--(d), respectively.} Before a turning point, there can be regions of unstable solutions (red curve). The instability region along the lower branch in panel (a) is $f_0\in(1.1199,1.2755)$. In panel (c), the instability is $f_0\in(2.0667,2.6635)$ and $(2.8905,3.1902)$. In panel (d), it is $f_0\in(1.1515,1.1998)$. The upper branch is always unstable.}		
	\label{fig:diamond_stub_bifur}
\end{figure}

\begin{figure}[tbhp!]
	\centering
	\subfloat[Diamond Lattice: stable profile for \(f_0 = 1\)]{\includegraphics[scale=0.58]{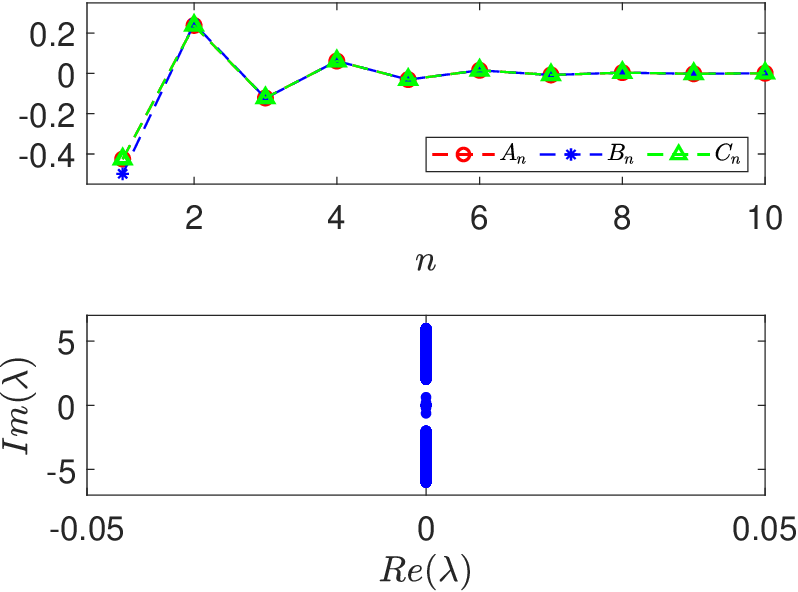}\label{subfig:prof_diamond_focusing_a}}\quad
	\subfloat[Diamond Lattice: unstable profile for \(f_0 = 1.2\)]{\includegraphics[scale=0.58]{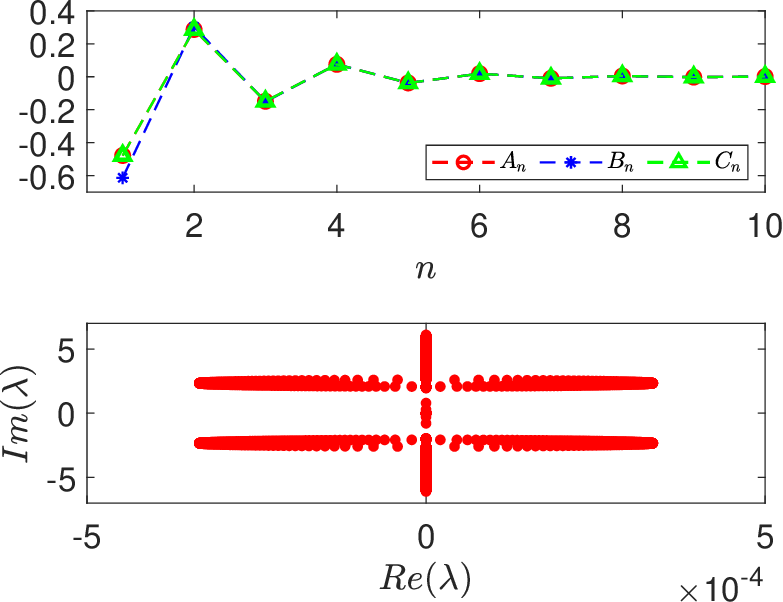}\label{subfig:prof_diamond_focusing_b}}\\
	\subfloat[Diamond Lattice: unstable profile for \(f_0 = 1.2\)]{\includegraphics[scale=0.58]{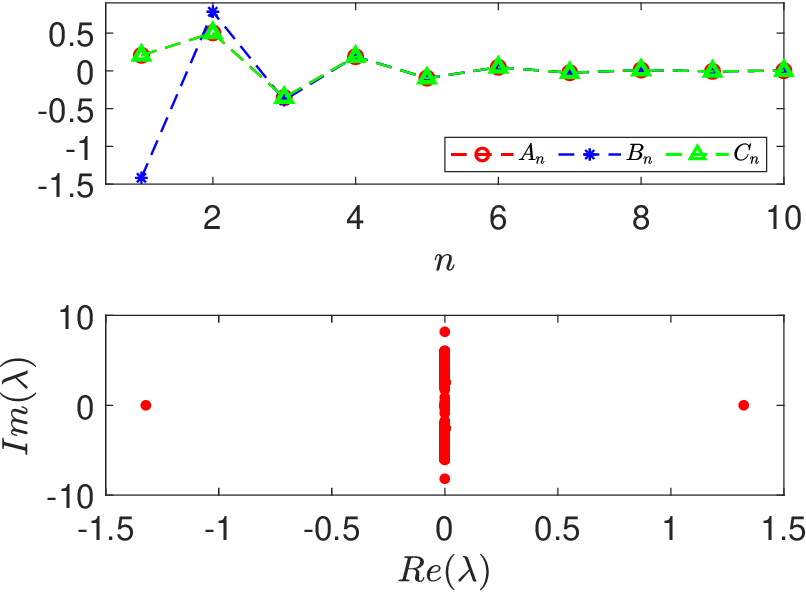}\label{subfig:prof_diamond_focusing_c}}\quad
	\subfloat[Stub Lattice: stable profile for \(f_0 = 1.5\)]{\includegraphics[scale=0.58]{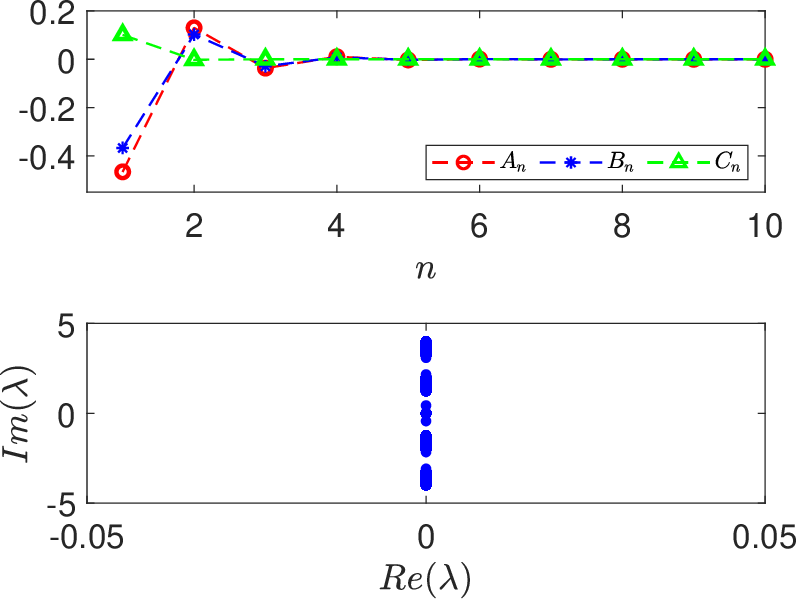}\label{subfig:prof_stub_focusing_d}}\\
	\subfloat[Stub Lattice: unstable profile for \(f_0 = 2.4\)]{\includegraphics[scale=0.58]{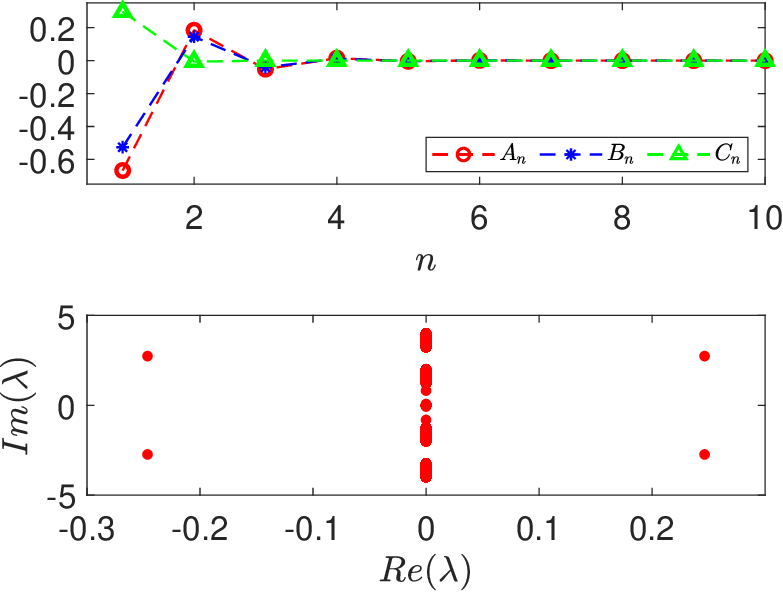}\label{subfig:prof_stub_focusing_e}}\quad	
	\subfloat[Stub Lattice: unstable profile for \(f_0 = 3.1\)]{\includegraphics[scale=0.58]{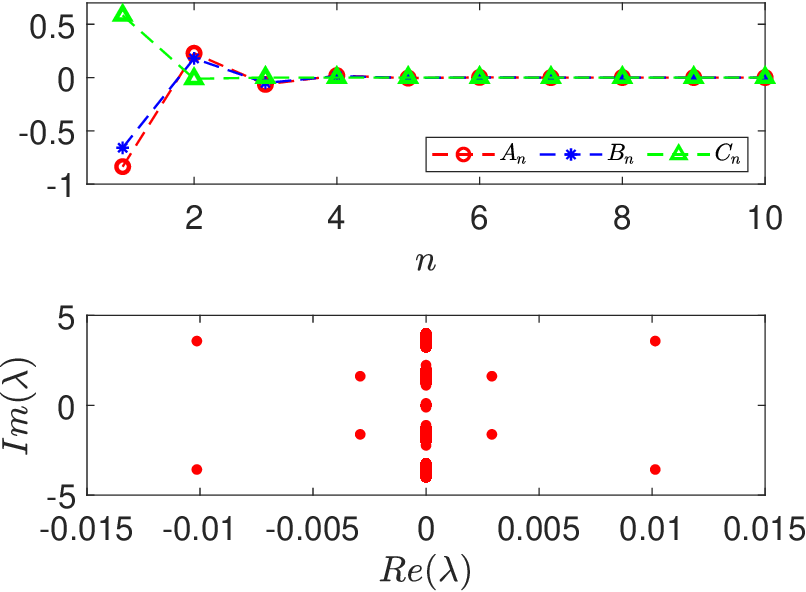}\label{subfig:prof_stub_focusing_f}}\\	
	\caption{Localized mode profiles in the focusing lattices for various \(f_0\) indicated in Fig.\ \ref{subfig:bifur_diamond_focusing} and \ref{subfig:bifur_stub_focusing}.} %The stable and unstable solutions are taken from the lower and upper branches in Figs.\ \ref{subfig:bifur_diamond_focusing} and \ref{subfig:bifur_stub_focusing}.}		
\label{fig:fig:diamond_stub_prof}
\end{figure}

Because of the edge drive \eqref{B0}, there will be an evanescent wave, i.e., localized mode, when the drive frequency $\Omega$ is not in the dispersive band. It is then essential to study the continuation of the decaying wave as we vary, e.g., the edge drive amplitude  \(f_0\) as a bifurcation parameter. To obtain bifurcation diagrams, we solve the time-independent equations \eqref{eq:supra_TI} for, let us say, \((\tilde{A}_n, \tilde{B}_n, \tilde{C}_n)\). The stability of the obtained solution is determined by introducing perturbations as follows:	
\begin{equation}
\begin{array}{rcl}
	A_n &=& \left(\tilde{A}_n + \epsilon_1 (\bar{v}_n^{(1)}(t) + i \bar{w}_n^{(1)}(t))\right)e^{-i\Omega t}, \\
	B_n &=& \left(\tilde{B}_n + \epsilon_2 (\bar{v}_n^{(2)}(t) + i \bar{w}_n^{(2)}(t))\right)e^{-i\Omega t}, \\
	C_n &=& \left(\tilde{C}_n + \epsilon_3 (\bar{v}_n^{(3)}(t) + i \bar{w}_n^{(3)}(t))\right)e^{-i\Omega t}.
\end{array}
\end{equation}
Substituting this into the governing equation \eqref{eq:supra_ori}, linearizing around \(\epsilon_1 = \epsilon_2 = \epsilon_3 = 0\), and separating real and imaginary parts, we obtain:
\begin{equation}\label{ev}
\dot{\mathbf{X}} = J \mathbf{X},
\end{equation}
where \(\mathbf{X} = (\tilde{v}_n^{(1)}, \tilde{w}_n^{(1)}, \tilde{v}_n^{(2)}, \tilde{w}_n^{(2)}, \tilde{v}_n^{(3)}, \tilde{w}_n^{(3)})^\text{T}\) and \(J\) is the Jacobian matrix given in terms of the operators \(\Delta_f\) and \(\Delta_b\), \(\gamma\), and coupling parameters, where 
\begin{equation}
J = \left(\begin{array}{cccccc}
	0&J_{12}&0&J_{14}&0&c_3\\
	J_{21}&0&J_{23}&0&-c_3&0\\
	0&J_{32}&0& J_{34}&0&J_{36}\\
	J_{41}&0&J_{43}&0&J_{45}&0\\
	0&c_3&0&J_{54}&0&J_{56}\\
	-c_3&0&J_{63}&0&J_{65}&0
\end{array}\right),
\end{equation}
\begin{equation*}
% \begin{array}{rcl}
	% 	J_{12}&=&\gamma \tilde{A}_n^2-2c_1-c_3-\Omega,\\
	% 	J_{14}&=&c_1\left(\Delta_{b}+I\right),\\
	% 	J_{21}&=&-3\gamma \tilde{A}_n^2+2c_1+c_3+\Omega,\\
	% 	J_{23}&=&-c_1\left(\Delta_{b}+I\right),\\
	% 	J_{32}&=&c_1\left(\Delta_{f}+I\right),\\
	% 	J_{34}&=&\gamma \tilde{B}_n^2-2c_1-2c_2-\Omega,\\
	% 	J_{36}&=&c_2\left(\Delta_{f}+I\right),\\
	% 	J_{41}&=&-c_1\left(\Delta_{f}+I\right),\\
	% 	J_{43}&=&-3\gamma \tilde{B}_n^2+2c_1+2c_2+\Omega,\\
	% 	J_{45}&=&-c_2\left(\Delta_{f}+I\right),\\
	% 	J_{54}&=&c_2\left(\Delta_{b}+I\right),\\
	% 	J_{56}&=&\gamma \tilde{C}_n^2-2c_2-c_3-\Omega,\\
	% 	J_{63}&=&-c_2\left(\Delta_{b}+I\right),\\
	% 	J_{65}&=&-3\gamma \tilde{C}_n^2+2c_2+c_3+\Omega.
	% \end{array}
\begin{array}{rclrcl}
	J_{12}&=&\gamma \tilde{A}_n^2-2c_1-c_3-\Omega,&
	J_{14}&=&c_1\left(\Delta_{b}+I\right),\\
	J_{21}&=&-3\gamma \tilde{A}_n^2+2c_1+c_3+\Omega,&
	J_{23}&=&-c_1\left(\Delta_{b}+I\right),\\
	J_{32}&=&c_1\left(\Delta_{f}+I\right),&
	J_{34}&=&\gamma \tilde{B}_n^2-2c_1-2c_2-\Omega,\\
	J_{36}&=&c_2\left(\Delta_{f}+I\right),&
	J_{41}&=&-c_1\left(\Delta_{f}+I\right),\\
	J_{43}&=&-3\gamma \tilde{B}_n^2+2c_1+2c_2+\Omega,&
	J_{45}&=&-c_2\left(\Delta_{f}+I\right),\\
	J_{54}&=&c_2\left(\Delta_{b}+I\right),&
	J_{56}&=&\gamma \tilde{C}_n^2-2c_2-c_3-\Omega,\\
	J_{63}&=&-c_2\left(\Delta_{b}+I\right),&
	J_{65}&=&-3\gamma \tilde{C}_n^2+2c_2+c_3+\Omega.
\end{array}
\end{equation*}
Here, \(I\) is the identity matrix.

Let \(\lambda\) denote the spectrum of the linearization operator \(J\) with the corresponding eigenvector $\mathbf{X}_v$. Along the special direction $\mathbf{X}_v$, the linear equation 
\eqref{ev} becomes $\dot{\mathbf{X}}_v=\lambda\mathbf{X}_v$, that has the solution $\mathbf{X}_v\sim e^{\lambda t}$. From this, it is clear that an equilibrium solution \((\tilde{A}_n, \tilde{B}_n, \tilde{C}_n)\) is said to be {linearly stable} if the perturbation $\mathbf{X}_v$ decays in time, i.e., \(\text{Re}(\lambda) \leq 0\) for \emph{all} eigenvalues of \(J\), and on the other hand, it is {unstable} if \emph{at least} one eigenvalue satisfies \(\text{Re}(\lambda) > 0\). {The imaginary part of $\lambda$, i.e., \(\text{Im}(\lambda)\) determines the oscillation frequency of the perturbation around the equilibrium. When \(\text{Re}(\lambda) = 0\), the perturbation exhibits oscillatory behavior at the frequency \(\text{Im}(\lambda)\), without growing or decaying over time.
} 
We compute bifurcation diagrams and plot the results as the norm of the localized modes, defined as $||\square_n|| = \left(\sum_n \left|\square_n\right|^2\right)^{\frac{1}{2}}$ against the bifurcation parameter \(f_0\), which represents the driving amplitude at the boundary. We also calculate the corresponding eigenvalues. The diagrams are color-coded, with blue and red lines indicating linearly stable and unstable branches.

We solve Eq.~\eqref{eq:supra_TI} for localized modes with parameter values at the flat band conditions. The drive frequency is taken to be precisely the same as the flat band. We plot the bifurcation diagrams in Fig.~\ref{fig:diamond_stub_bifur} for the diamond and stub lattices under focusing and defocusing conditions. We obtain that all panels have turning points. Beyond the points, there is no evanescent wave; we expect supratransmission instead. For the focusing diamond and stub lattices (Figs.~\ref{subfig:bifur_diamond_focusing} and \ref{subfig:bifur_stub_focusing}), the bifurcation diagrams show a stable branch that becomes unstable before the turning point. The stub lattice has wider existence and stability regions than the diamond lattice, presumably due to the limited connectivity. However, when the nonlinearity is defocusing (Figs.~\ref{subfig:bifur_diamond_defocusing} and \ref{subfig:bifur_stub_defocusing}), both lattices have relatively the same interval of existence. The lower branch is always stable in the diamond lattice.

\begin{figure}[tbph!]
\centering
\subfloat[Diamond Lattice Spectrum Focusing (Real)]{\includegraphics[scale=0.3]{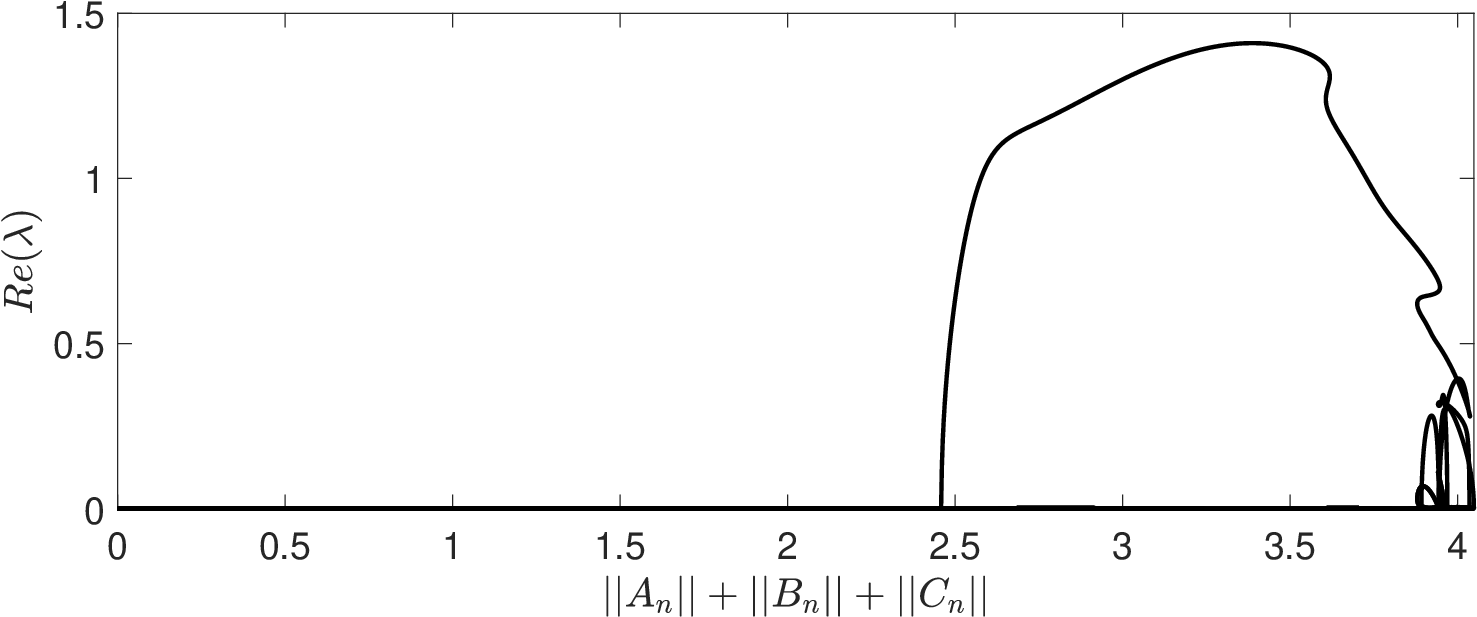}\label{subfig:spec_diamond_focusing_real}}\qquad
\subfloat[Diamond Lattice Spectrum Focusing (Imaginary)]{\includegraphics[scale=0.3]{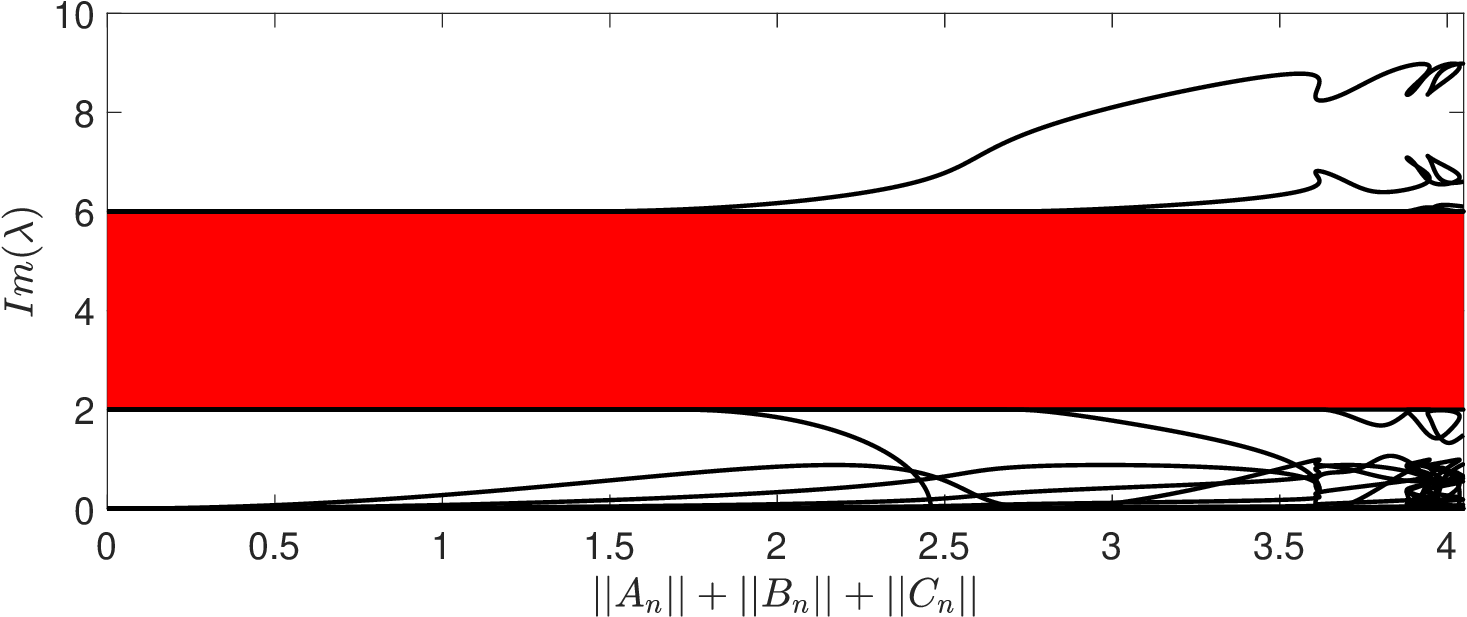}\label{subfig:spec_diamond_focusing_imag}}\\
\subfloat[Diamond Lattice Spectrum Defocusing (Real)]{\includegraphics[scale=0.3]{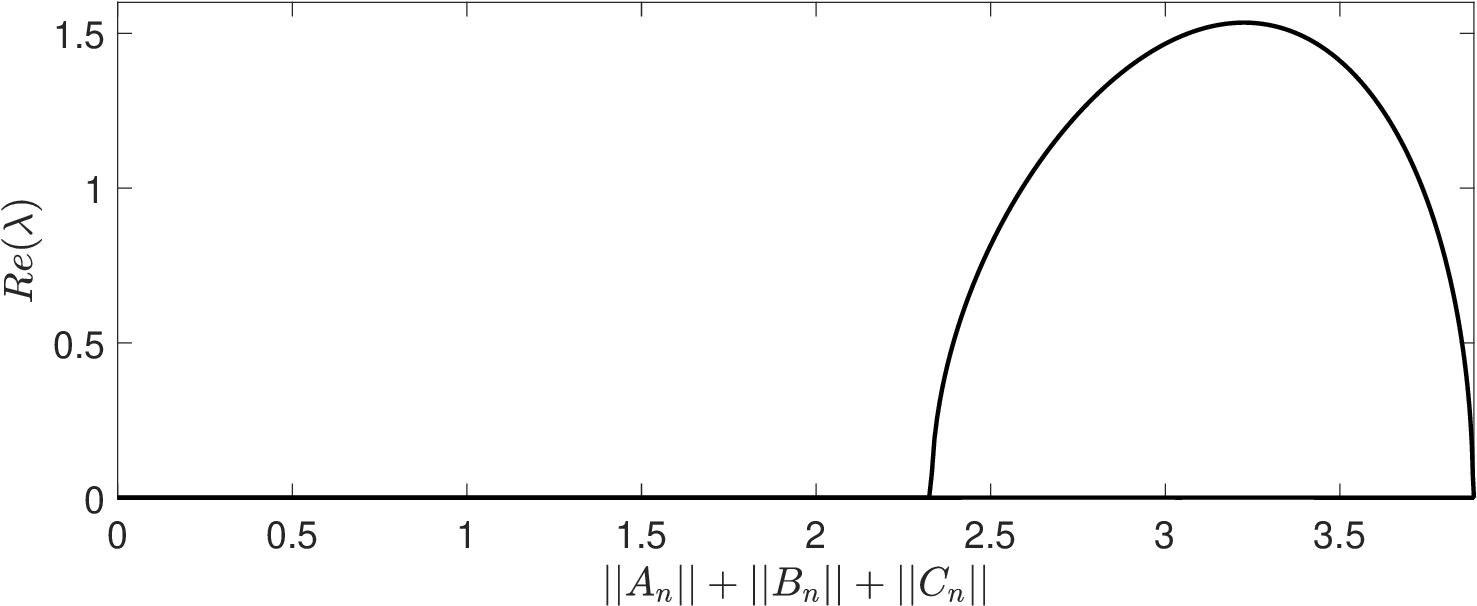}\label{subfig:spec_diamond_defocusing_real}}\qquad
\subfloat[Diamond Lattice Spectrum Defocusing (Imaginary)]{\includegraphics[scale=0.3]{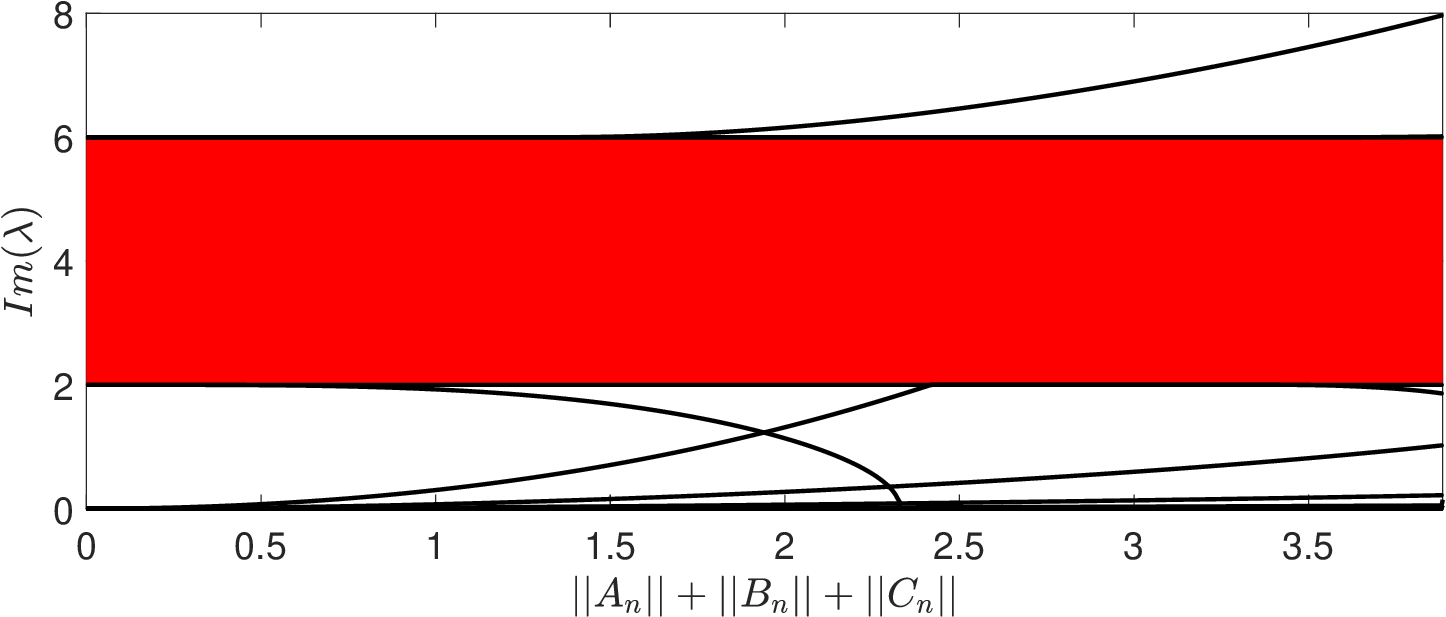}\label{subfig:spec_diamond_defocusing_imag}}\\
\subfloat[Stub Lattice Spectrum Focusing (Real)]{\includegraphics[scale=0.3]{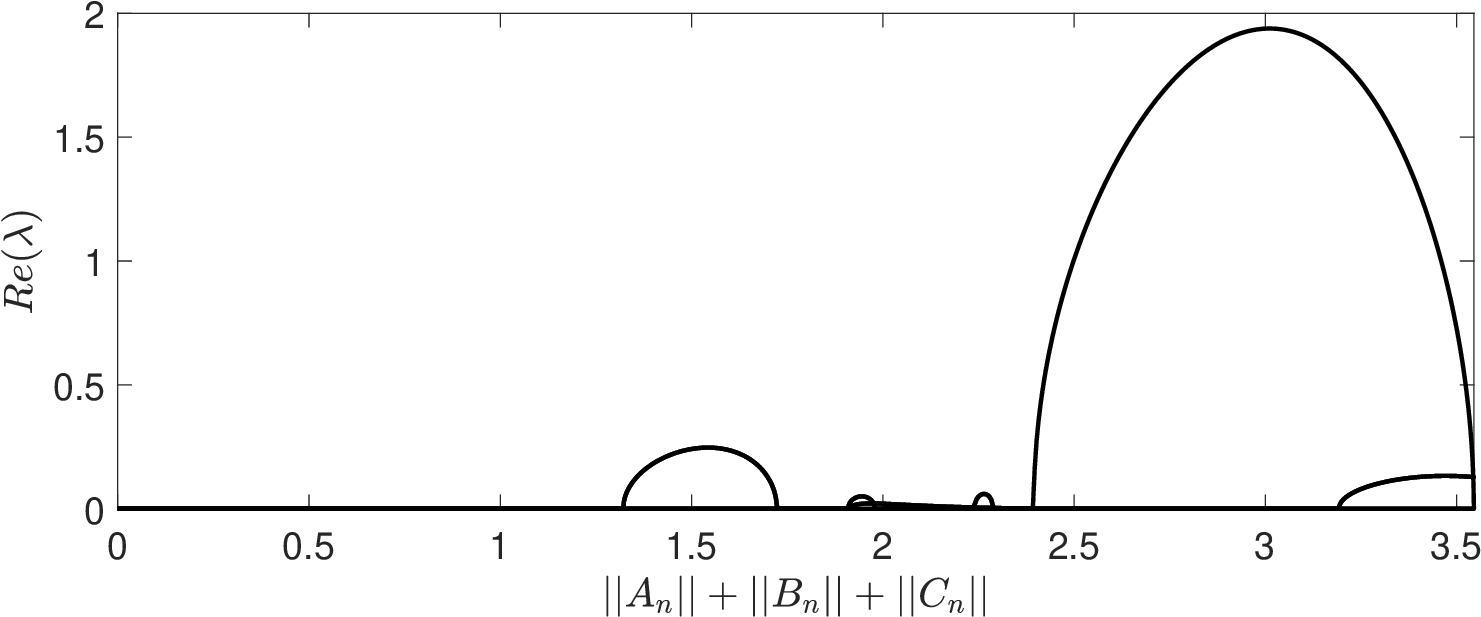}\label{subfig:spec_stub_focusing_real}}\qquad
\subfloat[Stub Lattice Spectrum Focusing (Imaginary)]{\includegraphics[scale=0.3]{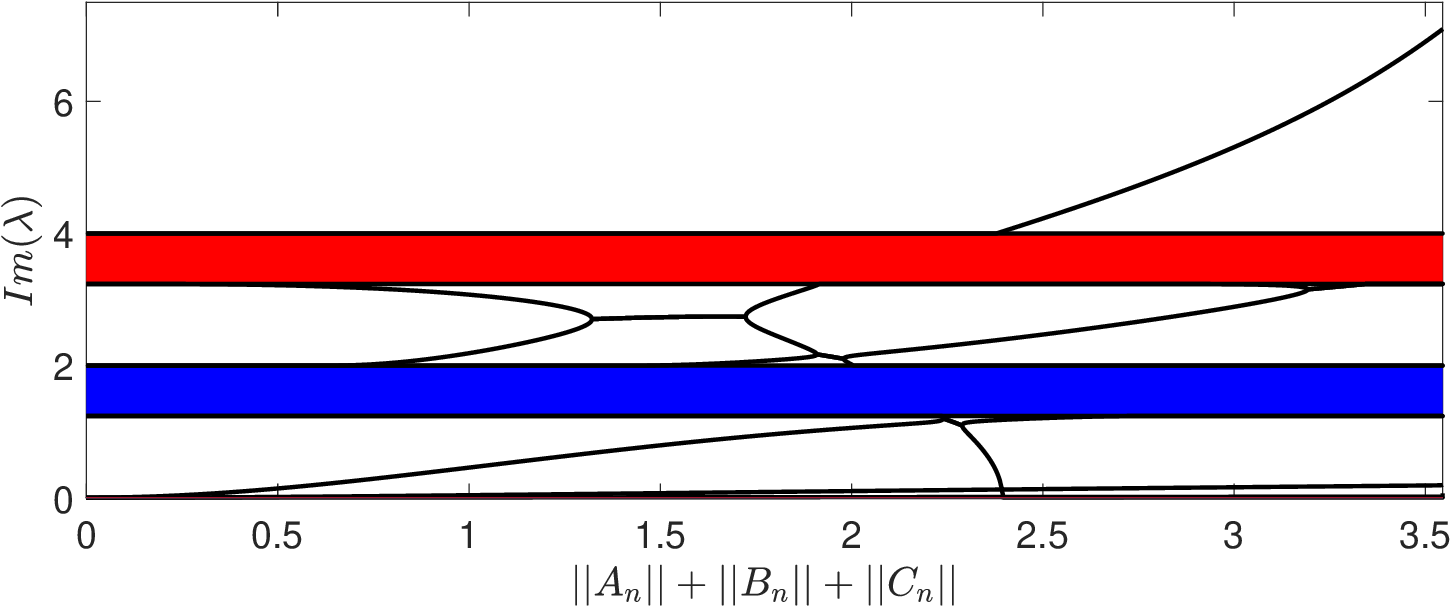}\label{subfig:spec_stub_focusing_imag}}\\
\subfloat[Stub Lattice Spectrum Defocusing (Real)]{\includegraphics[scale=0.3]{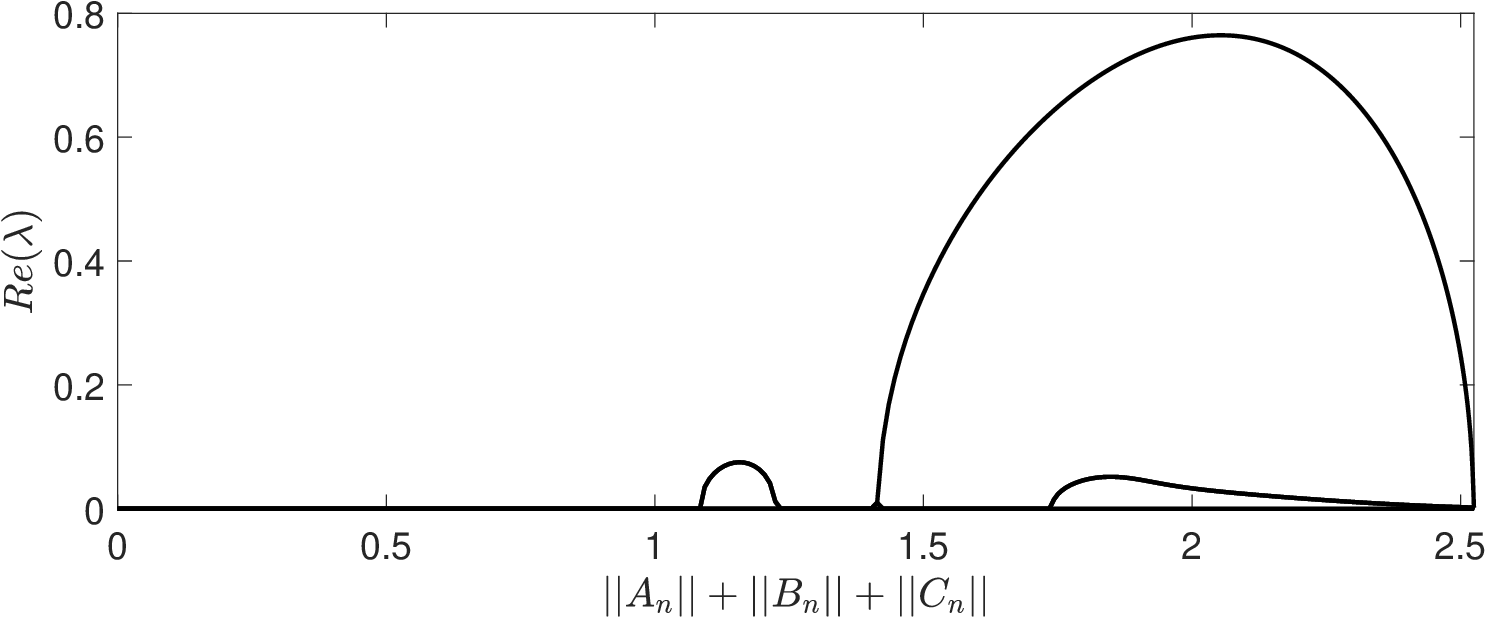}\label{subfig:spec_stub_defocusing_real}}\qquad
\subfloat[Stub Lattice Spectrum Defocusing (Imaginary)]{\includegraphics[scale=0.3]{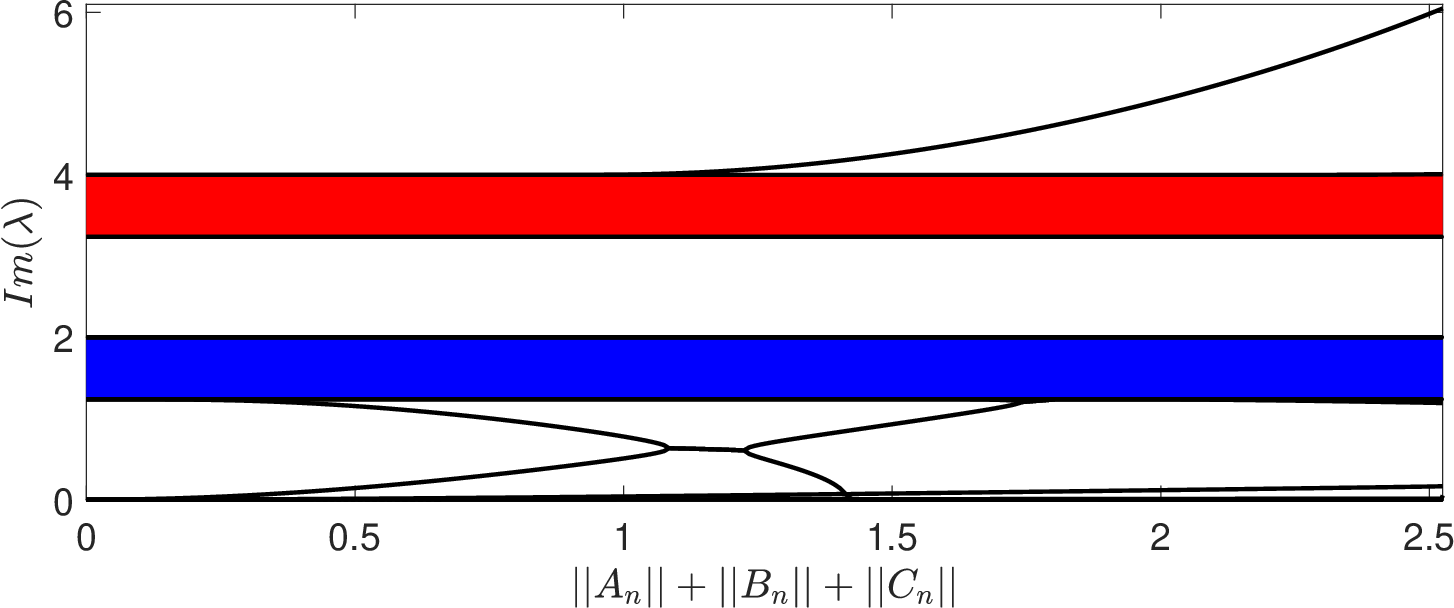}\label{subfig:spec_stub_defocusing_imag}}
\caption{Spectra of diamond and stub lattices under focusing and defocusing conditions. Real and imaginary parts of the spectra are shown for both lattice types, highlighting the stability and dynamical properties of localized modes. The continuous spectrum reflects the lattice's inherent translational symmetry, influencing wave propagation and stability.}
\label{fig:diamond_stub_spectrum}
\end{figure}

In Fig.~\ref{fig:fig:diamond_stub_prof}, we plot several stable and unstable localized mode profiles along the bifurcation diagrams, showing that the instability is caused by pairs of real eigenvalues (i.e., exponential instability) or complex ones (i.e., oscillatory instability). We also plot in Fig.\ \ref{fig:diamond_stub_spectrum} the spectrum along the bifurcation diagrams in Fig.\ \ref{fig:diamond_stub_bifur}. %We can note that there are also intervals where the instability is caused by complex eigenvalues with nonzero real parts, i.e., oscillatory instability. 
All unstable regions in the lower branch in Figs.\ \ref{fig:diamond_stub_bifur} are due to a quartet of eigenvalues. %Therefore, we obtain another difference caused by the lattice topology, where the instability in the diamond lattice is exponential, while in the stub, it is oscillatory. 

\section{Time-dependent dynamics}\label{sec:dependent}

\begin{figure}[tbhp!]
\centering
\subfloat[Diamond lattice for \(f_0 = 1.5\).]{\includegraphics[scale=0.35]{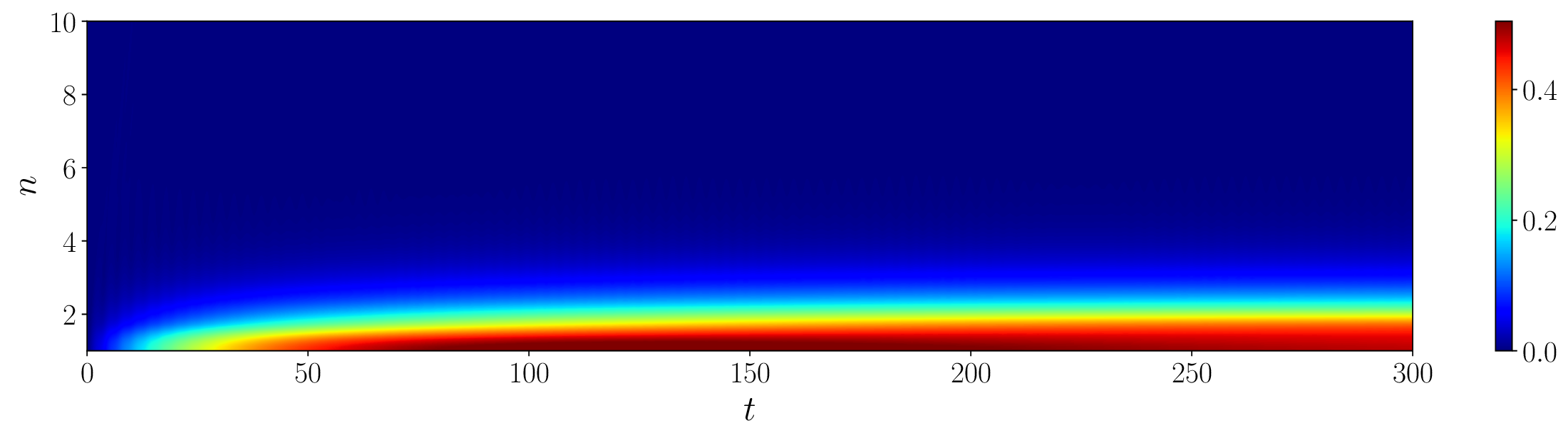}\label{subfig:Diamond_dynamics_notsupra}}\\
\subfloat[Diamond lattice for \(f_0 = 1.6\).]{\includegraphics[scale=0.35]{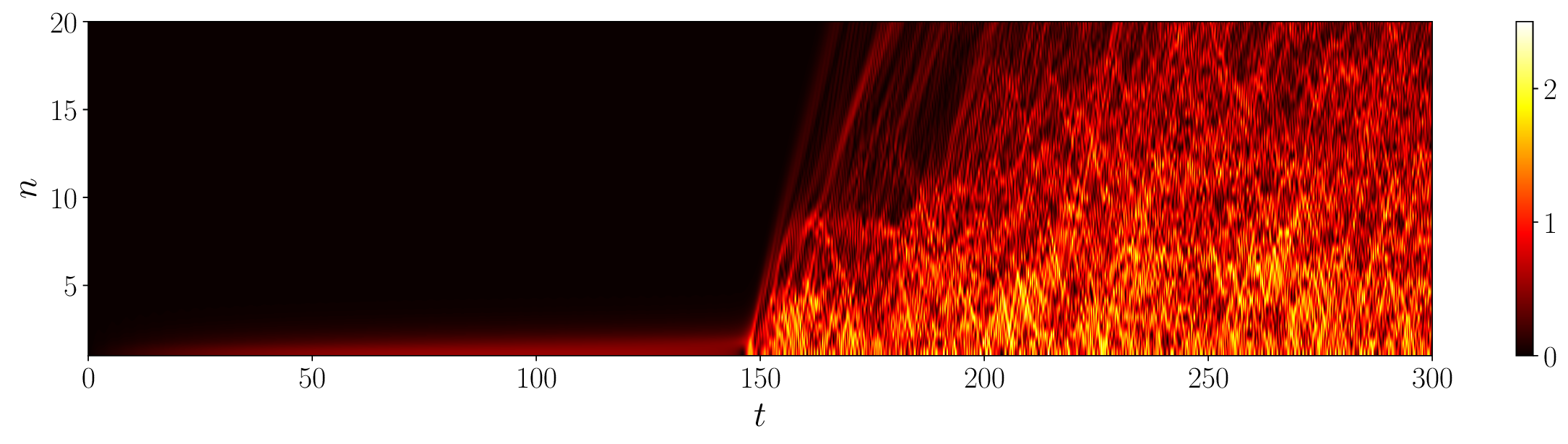}\label{subfig:Diamond_dynamics_supra}}\\
\subfloat[Stub lattice for \(f_0 = 2\).]{\includegraphics[scale=0.35]{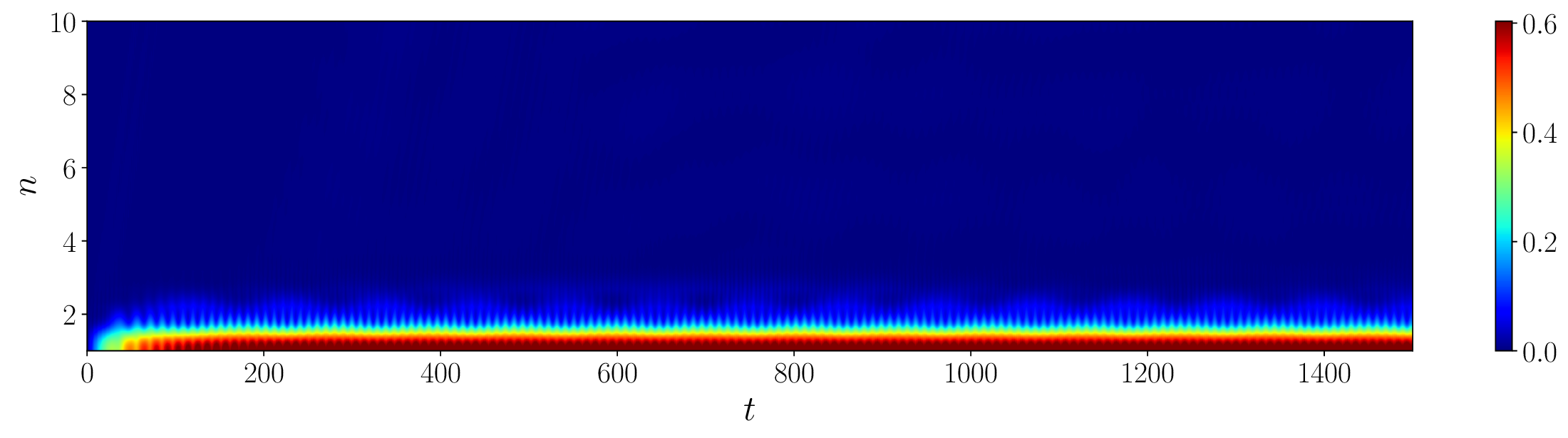}\label{subfig:Stub_dynamics_notsupra}}\\
\subfloat[Stub lattice for \(f_0 = 2.1\).]{\includegraphics[scale=0.35]{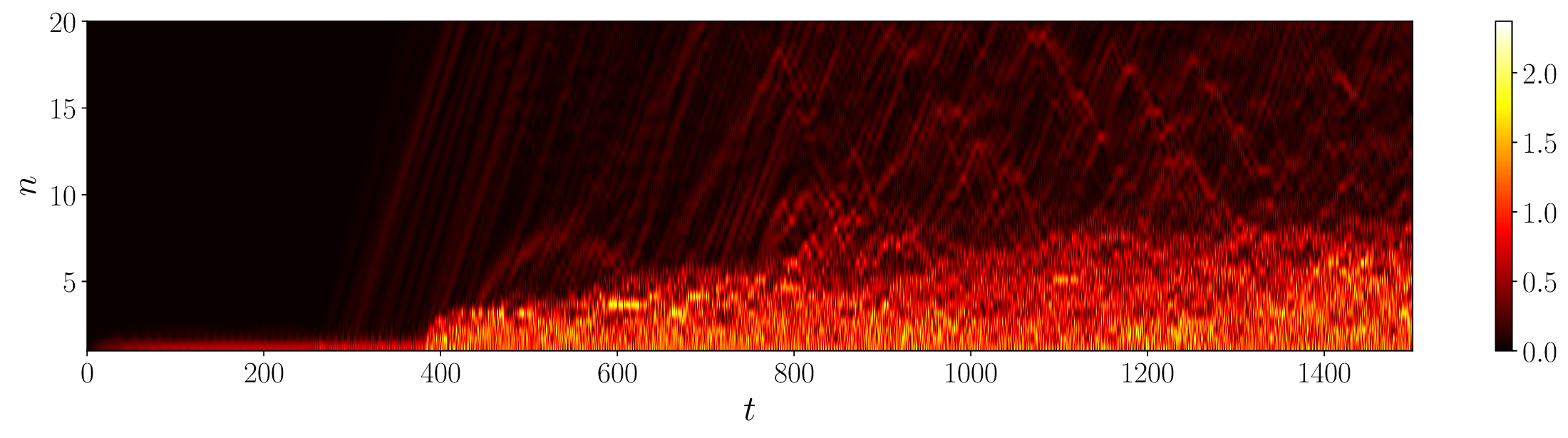}\label{subfig:Stub_dynamics_supra}}
\caption{Time dynamics of the diamond and stub lattices under focusing conditions for different driving amplitudes. Shown is the top view of the field $|A_n|$. $|B_n|$, and $|C_n|$ are not shown because the dynamics are similar. For the diamond lattice for \(f_0 = 1.5\) and the stub lattice for \(f_0 = 2\), the profile remains localized, reflecting the stability of localized modes. For the diamond lattice at \(f_0 = 1.6\) and the stub lattice at \(f_0 = 2.1\), the system surpasses the supratransmission threshold, resulting in energy propagation across the lattice.}
\label{fig:dynamics_diamond_stub}
\end{figure}

\begin{figure}[tbhp!]
\centering
\subfloat[Diamond Lattice Focusing.]{\includegraphics[scale=0.39]{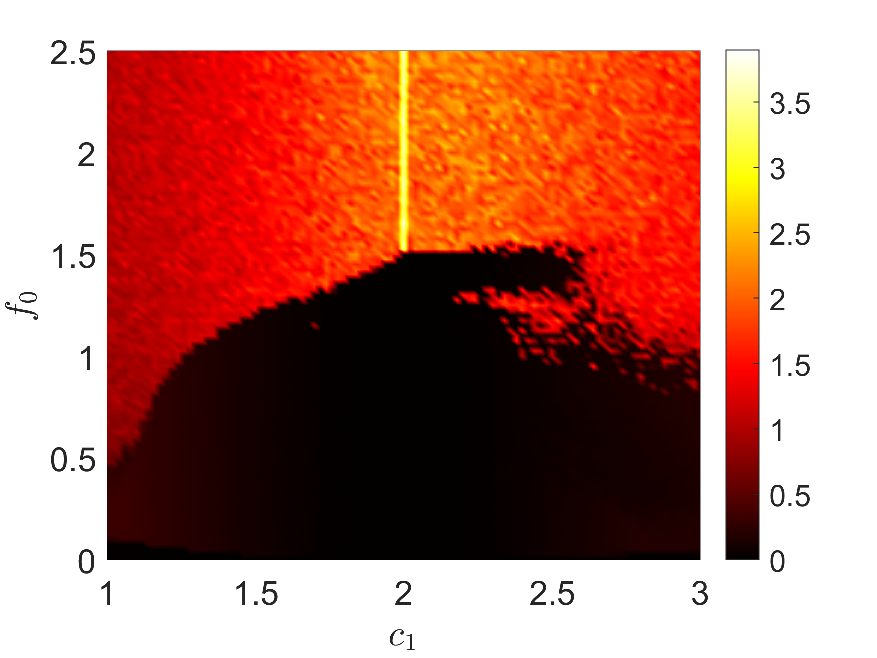}\label{subfig:Diamond_Focusing_power}}
\subfloat[Diamond Lattice Defocusing.]{\includegraphics[scale=0.39]{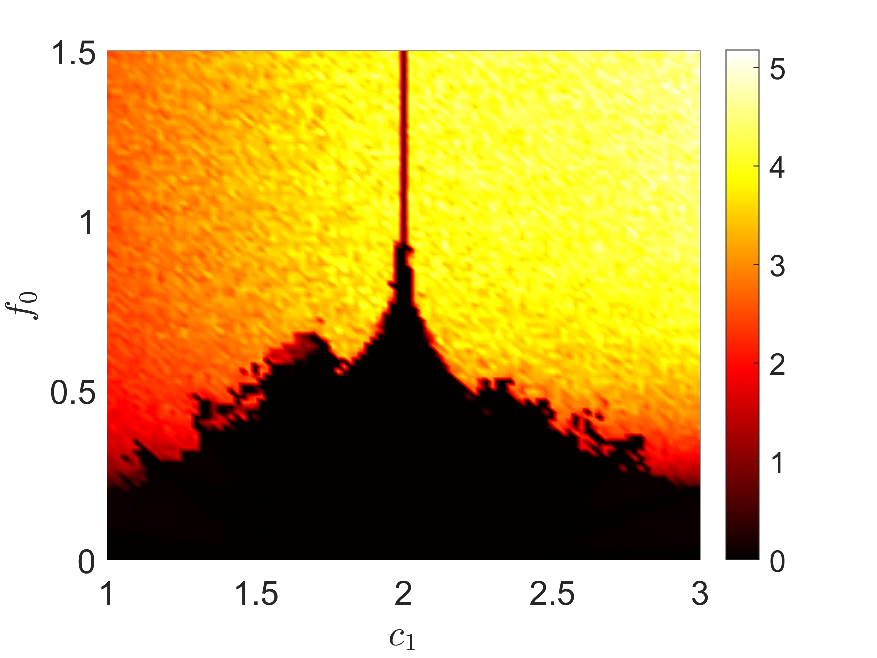}\label{subfig:Diamond_Defocusing_power}}
\subfloat[Diamond Lattice Linear.]{\includegraphics[scale=0.39]{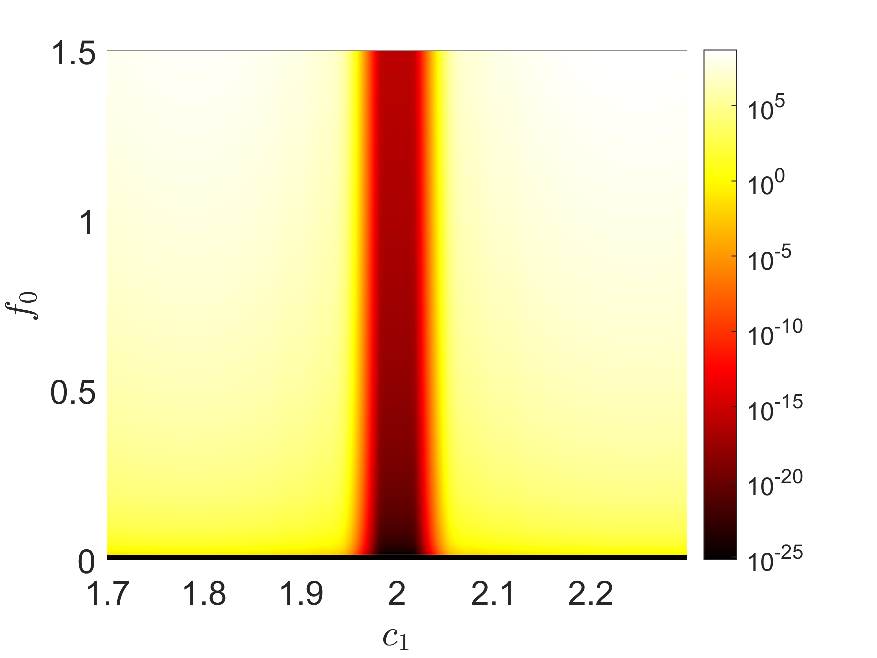}\label{subfig:Diamond_Linear_power}}\\
\subfloat[Diamond Lattice Focusing.]{\includegraphics[scale=0.39]{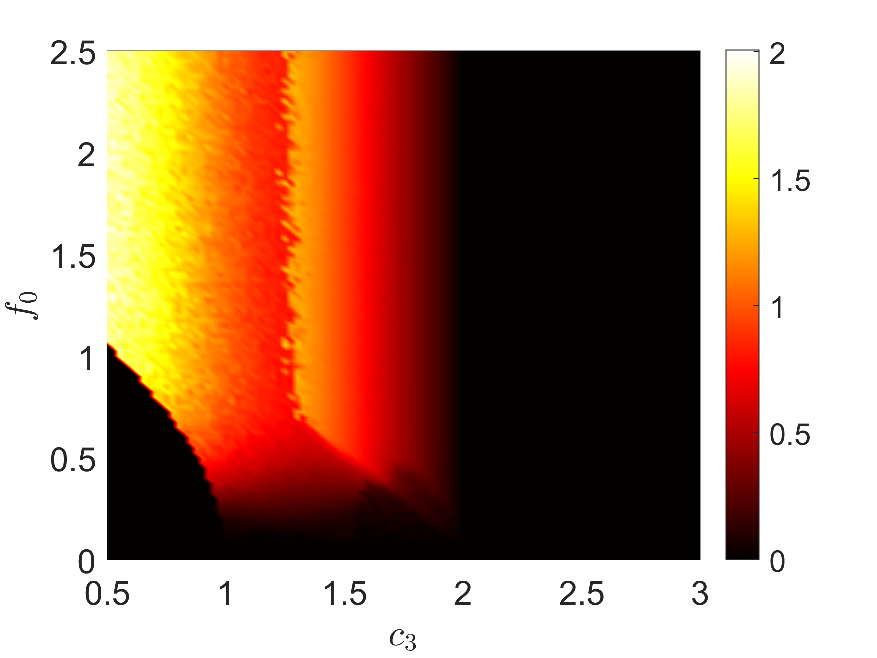}\label{subfig:Diamond_Focusing_power_inner}}
\subfloat[Diamond Lattice Defocusing.]{\includegraphics[scale=0.39]{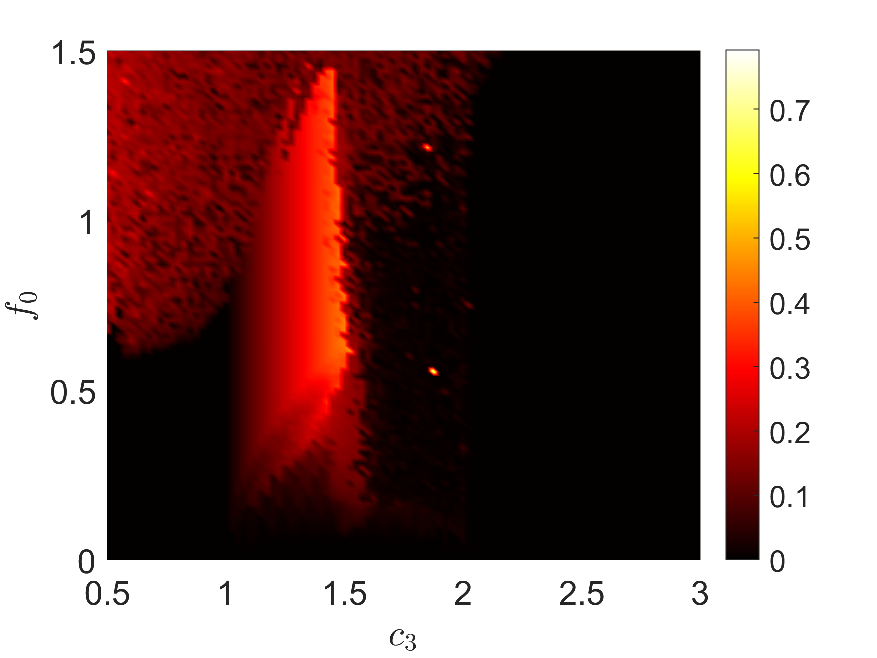}\label{subfig:Diamond_Defocusing_power_inner}}
\subfloat[Diamond Lattice Linear.]{\includegraphics[scale=0.39]{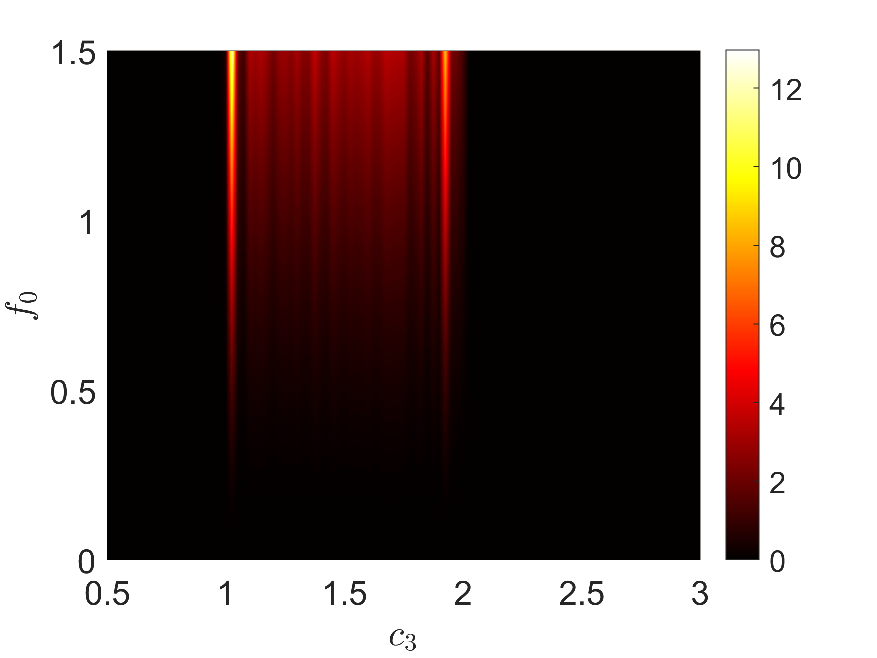}\label{subfig:Diamond_Linear_power_inner}}\\
\subfloat[Stub Lattice Focusing.]{\includegraphics[scale=0.39]{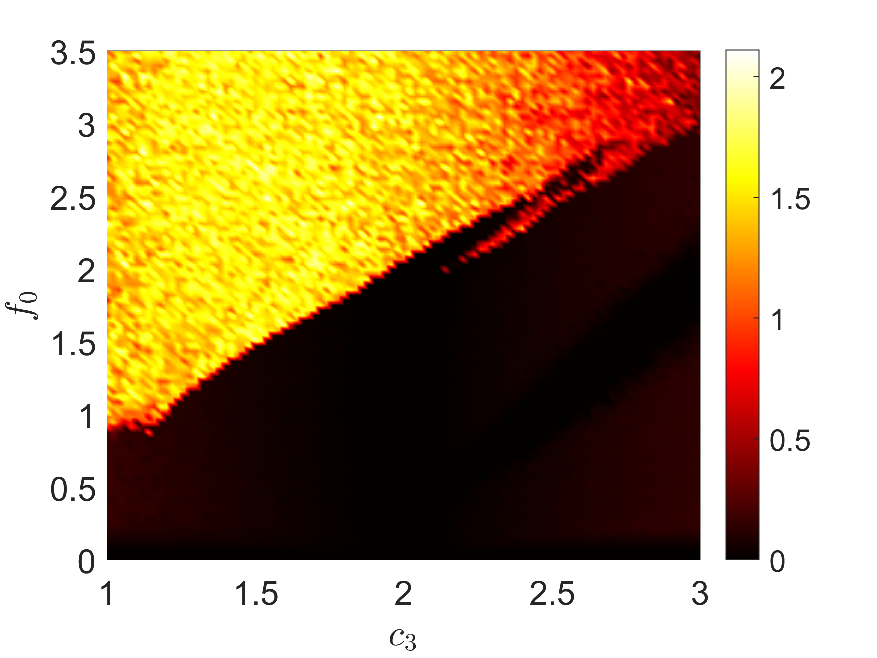}\label{subfig:Stub_Focusing_power}}	
\subfloat[Stub Lattice Defocusing.]{\includegraphics[scale=0.39]{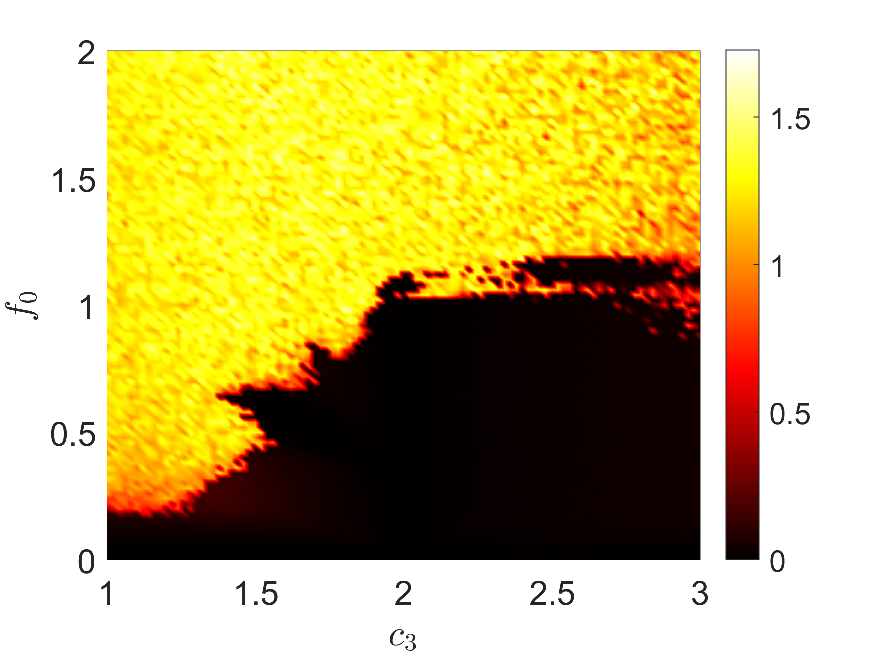}\label{subfig:Stub_Defocusing_power}}	
\subfloat[Stub Lattice Linear.]{\includegraphics[scale=0.39]{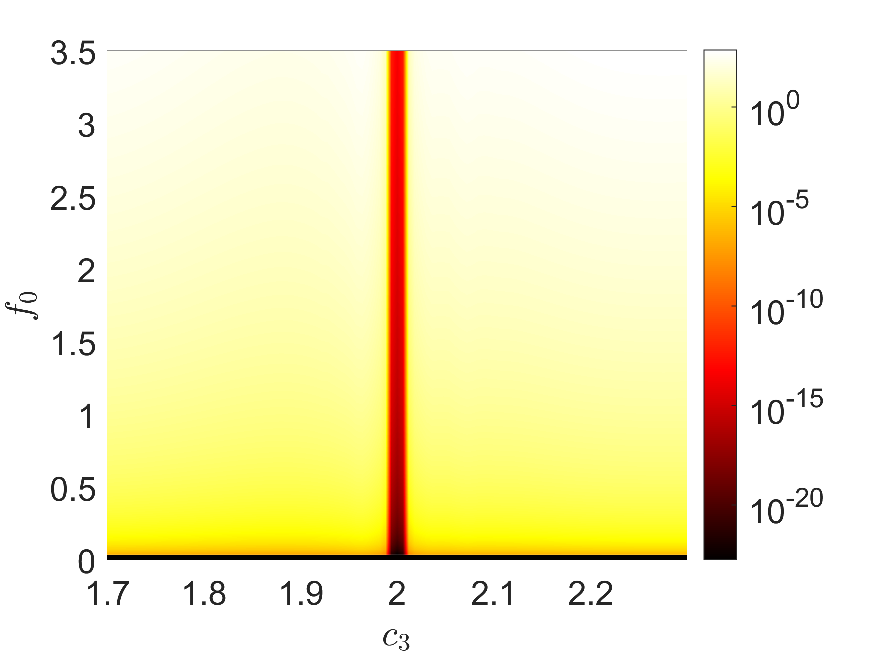}\label{subfig:Stub_Linear_power}}
\caption{
	{The transmission power for the diamond (a-c) (\(c_2=2,c_3=1\)), (d-f) (\(c_1=c_2=1\)), and (g-i) the stub (\(c_1=0,c_2=1\)) lattices under focusing, defocusing, and linear conditions. The surge of power at the flat band occurs at $f_0=1.51;0.945;2.06;1.13$  for panels (a), (b), (g), and (h), respectively.
		In panels (d) and (e), there exists a notable threshold amplitude for the surge of power when the flat band lies outside a dispersive band, i.e., $c_3<1$ or $c_3>2$. %Meanwhile, for $c_3\leq 1$, the surge appears only when $f_0$ reaches a sufficiently large value.
	}
	%    \textcolor{red}{(a)-(f) The transmission power for the diamond (\(c_2=2,c_3=1\)) and the stub (\(c_1=0,c_2=1\)) lattices under focusing, defocusing, and linear conditions. The surge of power at the flat band occurs at $f_0=1.51;0.945;2.06;1.13$  for panels (a), (b), (d), and (e), respectively. (g)-(i) The surge of power for the diamond lattice, where the flat band always exists ($c_1=c_2=1$), occurs when $c_3>1$, Meanwhile, for $c_3\leq 1$, the surge appears only when $f_0$ reaches a sufficiently large value.}
}
\label{fig:diamond_stub_power}
\end{figure}

%Our numerical results in Section \ref{sec:independent} show that there is a critical edge drive amplitude beyond which evanescent waves disappear. This is indicated by a turning point, i.e., saddle-node bifurcation. 
In the following, we will present simulations of the governing equations \eqref{eq:supra_ori}. We will study the implications of the steady states presented in Section \ref{sec:independent} on the time-dependent dynamics of the system. The model is integrated numerically using the fourth-order Runge-Kutta method with $h=10^{-2}$. In our computations, we use the number of sites $N=60$. We introduce increasing damping in the last $20$ sites to avoid reflections from the right end of the computation domain. 

Figure~\ref{fig:dynamics_diamond_stub} illustrates the time evolution of energy in the diamond and stub lattices under focusing conditions for two nearby values of driving amplitudes \(f_0\). These figures demonstrate how the driving amplitude governs the transition between energy confinement and supratransmission. For the diamond lattice, at \(f_0 = 1.5\) (Fig.~\ref{subfig:Diamond_dynamics_notsupra}), the dynamics show energy localized near the driving boundary with no significant propagation to the right end, indicating the absence of energy transmission. This behavior also shows that the localized mode is stable. However, when the driving amplitude increases to \(f_0 = 1.6\) (Fig.~\ref{subfig:Diamond_dynamics_supra}), the dynamics exhibit a pronounced characteristic shift. The wave propagates through the lattice, demonstrating supratransmission. This transition occurs because no localized mode exists at this drive value; see Fig.\ \ref{subfig:bifur_diamond_focusing}. In the stub lattice with focusing nonlinearity (Fig.~\ref{subfig:Stub_dynamics_notsupra}), a similar transition is observed at a higher driving amplitude, but below the turning point; see Fig.\ \ref{subfig:bifur_stub_focusing}. %possibly due to the lattice topology, i.e., reduced connectivity \(c_1=0\). 
We do not show the case of defocusing nonlinearity because the typical dynamics are the same as those of the focusing one. 

To understand the supratransmission systematically, we investigate the power flow injected by the drive into the array. To illustrate our findings, we calculate the time-averaged power at a specific site sufficiently far from the driving edge. For our case, we use $\langle |A_{25}|^2 + |B_{25}|^2 +|C_{25}|^2 \rangle$. The governing equations are integrated for $T_{\text{tr}} = 10,000$ time units during which the average power is computed. The results are displayed in Fig.~\ref{fig:diamond_stub_power} as a function of the driving force amplitude $f_0$ and the coupling constant $c_1$ (for the diamond lattice) and \(c_3\) (for the stub lattice). In all cases, the driving frequency $\Omega$ is set to be in the middle of the band that can become flat, i.e., the middle of the red-colored bands in Fig.\ \ref{fig:diamond_stub_freq}.

In all panels of Fig.~\ref{fig:diamond_stub_power}, a distinct feature emerges during supratransmission: an abrupt jump in the average power. This transition occurs at a threshold driving amplitude, \(f_0 = f_0^{\text{th}}\), which depends on the coupling parameters \(c_1\) (for the diamond lattice) and \(c_3\) (for the stub lattice). Below this threshold, power transmission to the cell at \(n = 25\) remains minimal. However, for \(f_0 > f_0^{\text{th}}\), the transmission stabilizes at consistently high levels, remaining nearly constant on average in the nonlinear lattice. We also plot the transmitted power of the linear lattices for completeness. There is no supratransmission in this case. However, a nearly flat band character is present, where power transmission is suppressed. Notably, the widths of the power suppression regions differ between the two lattices. When the red band is sufficiently broad, plane waves are transmitted regularly. 

We compare the threshold amplitudes \( f_0^{\text{th}} \) at the flat band in Fig.~\ref{fig:diamond_stub_power} with the critical drive amplitudes in Fig.~\ref{fig:diamond_stub_bifur}. For the diamond lattice, the threshold values---observed under both focusing and defocusing nonlinearities---align with the turning points where evanescent waves cease to exist. In contrast, the threshold drive amplitudes correspond to the onset of instability for the stub lattice. These findings demonstrate that the threshold amplitudes for supratransmission are not solely determined by either the turning point (at which evanescent waves vanish) or the instability onset (beyond which evanescent waves become unstable). Instead, these mechanisms play distinct roles depending on the lattice type.

Another significant observation is that the power surge persists even when the band is no longer flat. In diamond lattices, the flat band leaves a distinct signature, as evidenced by the characteristics of supratransmission illustrated by the vertical lines in panels (a) and (b) of Fig.\ \ref{fig:diamond_stub_power}. Beyond the threshold amplitude, the flat band lattice exhibits a distinct average transmitted power compared to neighboring regions of parameter values. In contrast, this signature is entirely absent in stub lattices.

{
In panels~(d) and~(e), we observe a variation in the threshold drive amplitude as we vary $c_3$ and the drive frequency approaches the dispersive band (see Fig.~\ref{fig:diamond_stub_freq}, panel~(b)), enters it, and subsequently exits it. When the driving frequency lies within the band, a finite critical drive amplitude still exists. Notably, the critical drive amplitude for supratransmission exhibits a discontinuous jump after the flat band exits the dispersive band. We have solved Eq.~\eqref{eq:supra_TI} for localized modes under these conditions, with the drive frequency set at the flat-band value, similar to Fig.~\ref{fig:diamond_stub_bifur}. For the focusing case in panel~(d), the bifurcation diagram shows no turning point, explaining the absence of supratransmission for $c_3>2$ at any value of $f$. For the defocusing case in panel~(e), the corresponding bifurcation diagram contains a turning point at a relatively small value of $f_0$. Although supratransmission occurs for $f>f_0$ in the region $c_3>2$, the transmitted power is significantly reduced. The cause of this behavior is currently unclear and is left for future investigation.}

\section{{Fourier spectra and topological states}}
%\begin{figure}[tbhp!]
%	\centering
%	\subfloat[Diamond]{\includegraphics[scale=0.5]{fft_diamond_f0_1_507}\label{subfig:fft_diamond_f0_1_507}}\quad
%	\subfloat[Stub]{\includegraphics[scale=0.5]{fft_stub_f0_2_0577}\label{subfig:fft_stub_f0_1_5}}\\
%	\subfloat[Diamond]{\includegraphics[scale=0.5]{fft_diamond_f0_1_508}\label{subfig:fft_diamond_f0_1_508}}\quad
%	\subfloat[Stub]{\includegraphics[scale=0.5]{fft_stub_f0_2_058}\label{subfig:fft_stub_f0_2_056}}\\
%	\subfloat[Diamond]{\includegraphics[scale=0.5]{fft_diamond_f0_1_51}\label{subfig:fft_diamond_f0_1_51}}\quad
%	\subfloat[Stub]{\includegraphics[scale=0.5]{fft_stub_f0_2_0582}\label{subfig:fft_stub_f0_2_057}}\\
%	\subfloat[Diamond]{\includegraphics[scale=0.5]{fft_diamond_f0_1_52}\label{subfig:fft_diamond_f0_1_52}}\quad
%	\subfloat[Stub]{\includegraphics[scale=0.5]{fft_stub_f0_2_06}\label{subfig:fft_stub_f0_2_058}}
%	\caption{}
%	\label{fig:fft_diamond_stub}
%\end{figure}
%\begin{figure}[tbhp!]
%	\centering
%	\subfloat[Diamond]{\includegraphics[scale=0.5]{fft_diamond_f0_0_7284}\label{subfig:fft_diamond_f0_1_507}}\quad
%	\subfloat[Diamond]{\includegraphics[scale=0.5]{fft_diamond_f0_0_7285}\label{subfig:fft_diamond_f0_1_508}}\quad
%	\subfloat[Diamond]{\includegraphics[scale=0.5]{fft_diamond_f0_0_729}\label{subfig:fft_diamond_f0_1_51}}\quad
%	\subfloat[Diamond]{\includegraphics[scale=0.5]{fft_diamond_f0_0_735}\label{subfig:fft_diamond_f0_1_52}}\quad
%	\caption{}
%	\label{fig:fft_diamond_stub}
%\end{figure}
\begin{figure}[tbhp!]
\centering
\subfloat[Diamond lattice]{\includegraphics[scale=0.38]{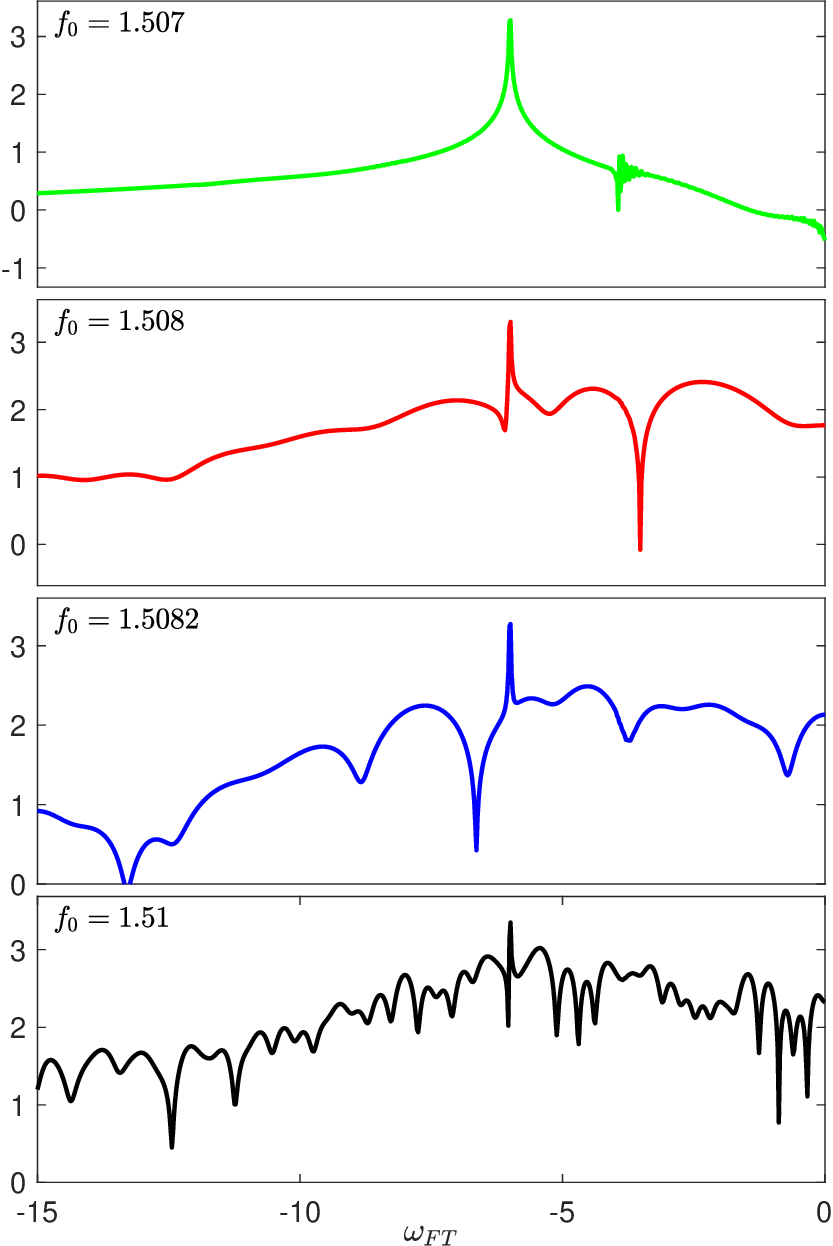}\label{subfig:fft_diamond_}}\,
%	\subfloat[Diamond lattice]{\includegraphics[scale=0.38]{fft_diamond_inner}\label{subfig:fft_diamond_inner}}\,
\subfloat[Stub lattice]{\includegraphics[scale=0.38]{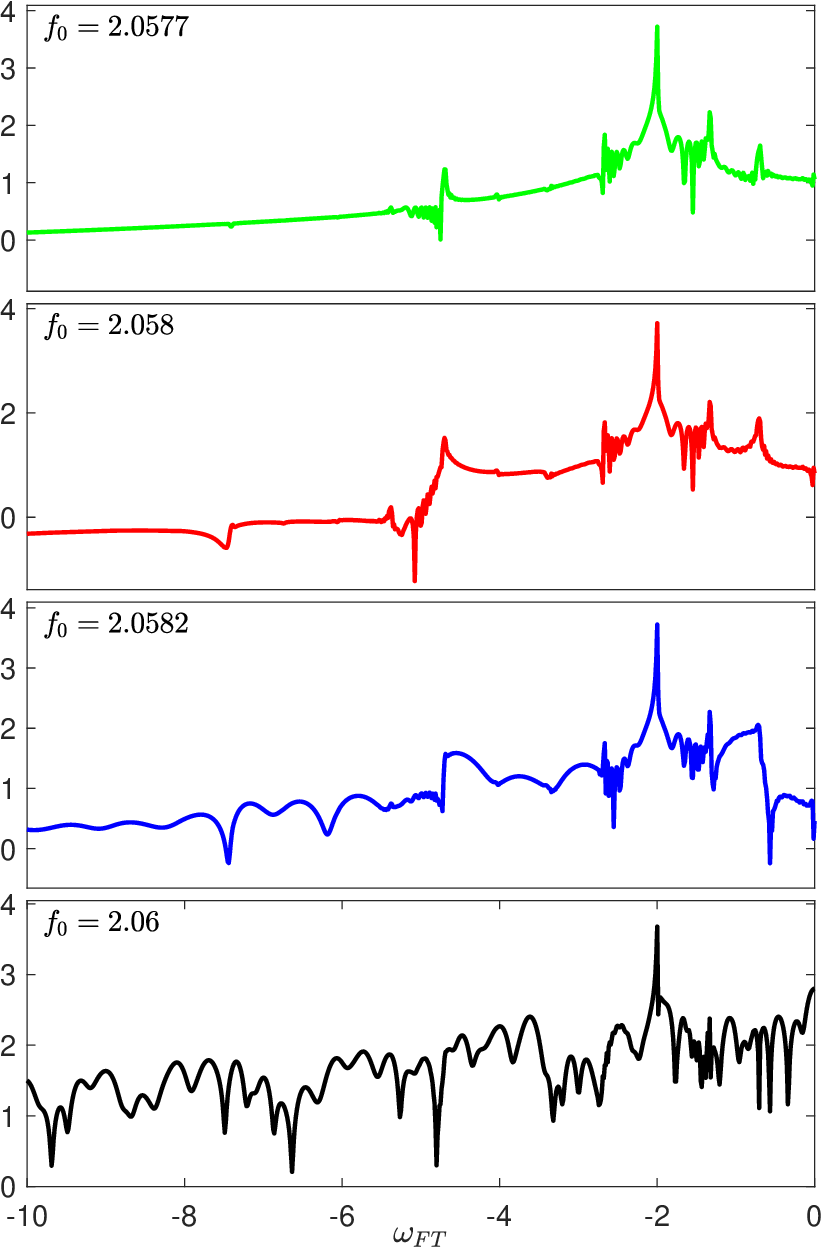}\label{subfig:fft_stub_}}
\caption{{The Fourier spectra of \(A_1(t)\) from the dynamics with parameters as in Fig.~\ref{fig:diamond_stub_power} are presented for several values of the driving amplitude \(f_0\). Panel (a) corresponds to the diamond lattice (\(c_1 = c_2 = 2, c_3 = 1\)), and panel (b) corresponds to the stub lattice (\(c_1 = 0, c_2 = 1, c_3 = 2\)). All spectra are displayed in semi-logarithmic plots, with \(\log_{10} \left[ \text{FT}\{A_1\} \right]\) on the vertical axis.
		%, the diamond lattice ($c_1=c_2=1,\ c_3=0.75$) in panel (b), 
	}
}
\label{fig:fft_diamond_stub}
\end{figure}
{
A closer inspection of Fig.~\ref{fig:diamond_stub_power} reveals a distinct feature in the dynamics of the evanescent waves between panels (a) and (c). In panel (c), a clear oscillation of the field is observed near the edge, prompting us to perform a Fourier spectral analysis of the dynamics. In Fig.~\ref{fig:fft_diamond_stub}, the decimal logarithms of the Fourier spectra for \( A_1(t) \), \( \log_{10}[{FT}\{A_1\}] \), are shown for the complete flat-band case as in Fig.~\ref{fig:diamond_stub_power}, for several values of the driving amplitude \( f_0 \).
}

{
For both lattices, when the driving amplitude \( f_0 \) is below the supratransmission threshold, the spectra are dominated by the driving frequency. In the diamond lattice case, for \( f_0 = 1.507 \), there is a strong peak at the driving frequency, accompanied by a secondary feature near \( \omega_{FT} \approx -4 \), which likely corresponds %to a topologically protected state with a frequency close 
to the lower edge of the top dispersive band (see Fig.~\ref{fig:diamond_stub_disper}). However, as \( f_0 \) is slightly increased, this secondary oscillation disappears.
}

{
This behavior differs in the stub lattice (right column). Although the lattice still shows a dominant peak at the driving frequency, a secondary peak appears to the left, corresponding to the frequency of the topologically protected state. Additionally, due to the nonlinearity, distinct subharmonics are observed. These additional peaks remain robust even as the drive amplitude increases.
}

{
Close to the threshold drive amplitude, the spectrum for the stub lattice maintains a somewhat equidistant structure, though some peaks broaden and weaken as nonlinear effects shift the resonant frequencies. Once \( f_0 \) surpasses the threshold, the spectrum undergoes an abrupt change: it becomes broadband and noisy, with significant power spread across the entire frequency window. Despite this broadband nature, the driving frequency remains visible, though less dominant. A similar feature in the spectra for supratransmission is also observed in the diamond lattice (left column). This abrupt transition from discrete subharmonics to a continuous spectrum is consistent with the onset of large-amplitude transmission, confirming that the supratransmission process is a genuine nonlinear transition.
}

\section{Conclusions} \label{sec:conc}
In conclusion, this study explores in-band supratransmission in nonlinear diamond and stub lattices, emphasizing the interplay between lattice topology, nonlinearity, and driving amplitude. Through numerical analysis, we demonstrate that supratransmission onset is governed by bifurcation thresholds tied to the driving amplitude \(f_0\). With its higher connectivity, the diamond lattice exhibits broader stability ranges and lower thresholds, facilitating greater energy transfer. In contrast, the stub lattice’s reduced connectivity increases its resistance to energy transfer, requiring higher \(f_0\) for supratransmission. Localized mode stability, spectral properties, and time dynamics reveal distinct energy confinement and propagation behaviors under focusing and defocusing conditions. These findings highlight the critical role of lattice topology and nonlinearity in shaping energy dynamics, which may offer valuable insights for applications in photonic crystals, optical lattices, and phononic systems.

{An important direction for future work is to explore the effects of disorder and lattice imperfections on supratransmission thresholds in flat-band lattices. Introducing disorder can break the perfect lattice symmetry that underpins the existence of compact localized modes and flat bands \cite{leykam2013flat,roy2020interplay,orito2021interplay,rivas2020seltrapping,leykam2017localization}. This may lead to partial delocalization, spectral broadening, or the emergence of mobility edges. However, we expect that the basic mechanism of supratransmission will still work in moderately disordered systems, though with different thresholds \cite{yamgoue2007noise,yousefzadeh2016supratransmission}. Understanding these effects is essential for assessing the feasibility of wave control in realistic physical systems, where imperfections and fabrication tolerances are inevitable.	}

%Moderate disorder on the sites might change the supratransmission thresholds and cause more localization. 

\section{Aknowledgements}
\textbf{RK} acknowledges the budget efficiency of all institutions in their collective non-involvement, while family generously ignored such policy.
%acknowledges Riset Utama PPMI FMIPA 2024 (617I/IT1.C02/KU/2024) and Riset ITB 2025. 
\textbf{HS} acknowledged support by Khalifa University through %a Faculty Start-Up Grant (No.\ 8474000351/FSU-2021-011), 
a Competitive Internal Research Awards Grant (No.\ 8474000413/CIRA-2021-065) and Research \& Innovation Grants (No.\ 8474000617/RIG-S-2023-031 and No.\ 8474000789/RIG-S-2024-070). 

\bibliographystyle{elsarticle-num}
\bibliography{references}
\end{document}